\documentclass[%
pre,
tightenlines,
showpacs,showkeys,
a4paper,12pt
]{revtex4}
\pdfoutput=1
\usepackage{dcolumn}
\usepackage{amssymb,amsmath,stmaryrd,array}

\usepackage{graphicx}
\usepackage{subfigure}

\renewcommand{\thesubfigure}{(\alph{subfigure})}

\makeatletter
\renewcommand{\@thesubfigure}{\thesubfigure\space}

\newcommand{\sca}[2]{\ensuremath{\bigl({#1},{#2}\bigr)}}

\newcommand{\cnj}[1]{{#1}^{\ast}}
\newcommand{\hcnj}[1]{{#1}^{+}}
\newcommand{\tcnj}[1]{{#1}^{T}}

\newcommand{\pdrs}[1]{\partial_{#1}}

\newcommand{\diag}{\mathop{\rm diag}\nolimits}
\newcommand{\sign}{\mathop{\rm sign}\nolimits}
\renewcommand{\Re}{\mathop{\rm Re}\nolimits}
\renewcommand{\Im}{\mathop{\rm Im}\nolimits}

\newcommand{\mum}{$\mu$m}

 \newcommand{\bs}[1]{\boldsymbol{#1}}
 \newcommand{\vc}[1]{\mathbf{#1}}
 \newcommand{\mvc}[1]{\mathbf{#1}}
 \newcommand{\uvc}[1]{\hat{\mathbf{#1}}}

\newcommand{\dd}{\mathrm{d}}

 \newcommand{\ee}{\mathrm{e}}

\newcommand{\inc}{\mathrm{inc}}
\newcommand{\refl}{\mathrm{refl}}
\newcommand{\trans}{\mathrm{tr}}
\newcommand{\vac}{\mathrm{vac}}
\newcommand{\med}{\mathrm{m}}
\newcommand{\ellpt}{\mathrm{ell}}

\makeatother

\begin{document}
\DeclareGraphicsExtensions{.jpg,.png,.pdf}
\title{Topological events in 
polarization resolved angular patterns of nematic liquid crystal cells
at varying ellipticity of incident wave} 

\author{Alexei~D.~Kiselev}
\email[Email address: ]{kiselev@iop.kiev.ua}

\author{Roman~G.~Vovk}
\email[Email address: ]{roman.vovk@gmail.com}

\affiliation{%
 Institute of Physics of National Academy of Sciences of Ukraine,
 prospekt Nauki 46,
 03028 Ky\"{\i}v, Ukraine} 

\date{\today}

\begin{abstract}
We study the angular structure of polarization of light transmitted 
through a nematic liquid crystal (NLC) cell 
by analyzing the polarization state as a function of the incidence
angles and the polarization of the incident wave. 
The polarization resolved angular (conoscopic) patterns 
emerging after the  NLC cell illuminated by the convergent light beam
are described in terms
of the polarization singularities such as C-points 
(points of circular polarization) and L-lines (lines of linear polarization). 
For the homeotropically aligned cell,
the Stokes polarimetry technique is used to
measure the polarization resolved conoscopic patterns 
at different values of the ellipticity of the incident light, $\epsilon_{\ellpt}^{(\inc)}$.
impinging onto the cell.
Using the exact analytical expressions for 
the transfer matrix we 
show that variations of the ellipticity, $\epsilon_{\ellpt}^{(\inc)}$,
induce transformations of the angular pattern
exhibiting the effect of \textit{avoided L-line crossings}
and characterized by topological events such as
\textit{creation and annihilation of the C-points}.
The predictions of the theory 
are found to be in good agreement with the experimental
results.
\end{abstract}

\pacs{%
45.25.Ja, 78.20.Fm, 42.70.Df, 42.25.Bs 
}
\keywords{%
polarization of light; nematic liquid crystal;  polarization singularities 
} 
 \maketitle

\section{Introduction}
\label{sec:intro}

Singularities that represent structurally stable topological defects
have long been known to play a pivotal role in condensed matter 
physics~\cite{Luben:bk:1995}. They are of particular importance
in determining the properties of systems with a broken continuous 
symmetry~\cite{Merm:rpm:1979,Michel:rpm:1980,Trebin:advph:1982,Klem:bk:1984}
(for a recent review see, e.g., Ref.~\cite{Kleman:rpm:2008}).
Topological methods of quantum field and gauge 
theories have been extensively used to classify the topological 
defects and to describe 
transformations of the
singularities in ordered media such as superfluids and 
liquid crystals~\cite{Volovik:rpm:1987,Monast:bk:1993,Schw:bk:1993}.
 
For electromagnetic vector fields,
it was originally recognized by 
Nye~\cite{Nye:prsl:1983a,Nye:prsl:1987,Nye:bk:1999}
that the so-called \textit{polarization singularities} are
the important elements characterizing geometry of 
the Stokes parameter fields.
In particular, 
the polarization singularities such as the \textit{C-points} 
[the points where the light wave is circularly polarized]
and the \textit{L-lines}
[the curves along which the polarization is linear] 
frequently emerge as the characteristic feature 
of certain polarization state distributions. 
Over the past two decades these singularities and related issues
have been the subject of numerous theoretical, experimental and numerical 
studies~\cite{Hajnal:prsl:1987a,Hajnal:prsl:1987b,Hajnal:prsl:1990,
Freund:pra:1994,Freund:optl:2002,Dennis:prsl:2000,Dennis:prsl:2001a,
Melnikov:jopb:2001,Freund:optcom:2002,Soskin:optcom:2002,
Mokhun:optl:2002,Dennis:optcom:2002}.

The theory of polarization singularities
has also been found to be a useful tool for studying
optical properties of anisotropic media. 
In Ref.~\cite{Dennis:prsl:2003}, 
it was applied to study the angular dependence 
of the polarization state of the electric displacement field
for plane wave eigenmodes in birefringent dichroic chiral
crystals.
This analysis was then generalized and extended to 
a more complicated case of bianisotropic media~\cite{Berry:prsl:2005}.

The experimental results and theoretical analysis presented
in Refs.~\cite{Dennis:prl:2005,Floss:optex:2006} 
deal with the unfolding of
a linearly polarized Laguerre-Gauss (LG$_{01}$) beam
with an on-axis vortex
on propagation through a birefringent crystal.
It was found that 
a complicated pattern of polarization singularities is formed
as a result of  the anisotropy induced symmetry breaking.

In our recent investigation
into the two-dimensional angular distributions of the
Stokes parameters describing 
the polarization structure behind the conoscopic 
images we performed a comprehensive theoretical analysis of the
polarization state of the light transmitted through 
nematic liquid crystal (NLC) cells
as a function of the incidence angles~\cite{Kis:arxiv:2006,Kis:jpcm:2007}.
This polarization structure~---~the so-called
 \textit{polarization resolved angular (conoscopic) pattern}~---~is 
represented by the field of polarization ellipses
and results from
the interference of four eigenmodes excited in NLC cells
by the plane waves with varying direction of incidence.
 
Note that NLCs are technologically important
materials where the optical anisotropy is determined by
the orientational structure which is sensitive to external
fields and, in restricted geometries, 
can also be influenced by changing the boundary 
conditions~\cite{Gennes:bk:1993,Yeh:1998,Chigr:1999}.
The conoscopy is widely used as 
an experimental technique  to characterize orientational structures in NLC
cells.

For example, this method was employed to detect biaxiality of
NLCs~\cite{Yu:prl:1980,Madsen:prl:2004} and
to measure the pretilt angle in uniaxial liquid crystal
cells~\cite{Komit:crt:1984,Brett:apo:2001}.
Orientational structures and helix unwinding process in
ferroelectric and antiferroelectric 
smectic liquid crystals were also studied by conoscopy
in Refs.~\cite{Gorec:jjap:1990,Elston:jjap:2002,Suwa:jjap:2003}.

In this paper we explore both theoretically and experimentally
the polarization resolved angular patterns
for convergent light beam impinging on
the homeotropically aligned NLC cell at varying polarization of the
incident wave.
The layout of the paper is as follows.


In Sec.~\ref{sec:experiment} we
give experimental details and
describe our setup employed to carry out measurements
using a suitably modified method of
the Stokes 
polarimetry~\cite{Soskin:optl:2003,Soskin:jopa:2004,Dennis:prl:2005,Kis:ujph:2007}.

The problem of light transmission through a uniformly
anisotropic NLC cell is considered
in Sec.~\ref{subsec:transm-problem}.
Using the $4\times 4$ matrix formalism
we deduce the general expressions
relating the evolution matrix to
the transmission and reflection matrices.
The analytical results are used for analysis
of the polarization resolved conoscopic patterns.
in Sec.~\ref{subsec:pol-resolv-pattern}

The patterns emerging after 
homeotropically aligned NLC cells
are treated in Sec.~\ref{subsec:ellipt-dynam}. 
The polarization ellipse fields 
are studied as a function of 
the ellipticity, $\epsilon_{\ellpt}^{(\inc)}$,
and the polarization azimuth, $\phi_p^{(\inc)}$,
characterizing the polarization state of the incident wave.

The analytical expressions
describing loci of the C-points and L-lines
along with the index formula
are used to examine rearrangements
of the polarization singularities
caused by variations of the polarization
parameters $\epsilon_{\ellpt}^{(\inc)}$ and
$\phi_p^{(\inc)}$.

We find that changing the azimuth 
results in rotation of the ellipse field
as a whole by the angle $\phi_p^{(\inc)}$,
whereas the transformations induced by the ellipticity
appear to be complicated
by the presence of bifurcations
leading to creation and annihilation of
the C-points. 
It is shown that  
the structure of the intersecting L-lines
formed at $\epsilon_{\ellpt}^{(\inc)}=0$
smoothly evolve into a family of 
the concentric L-circles at $|\epsilon_{\ellpt}^{(\inc)}|=1$.
At small values of the ellipticity,
the effect of avoided L-line crossings
is found to be
an important feature of this transition.
 
Experimentally measured and computed
fields of the polarization ellipses are presented in
Sec.~\ref{sec:results} along with 
discussion and concluding remarks.
Mathematical details on the evolution operator
of uniformly anisotropic media and related issues
are relegated to Appendix~\ref{sec:evolution_oper}.

\begin{figure*}[!tbh]
\centering
   \resizebox{155mm}{!}{\includegraphics*{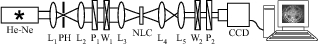}}
\caption{%
Experimental setup: 
He-Ne is the laser; 
L$_1$, L$_2$ and L$_5$ are the collimating lenses;
L$_3$ and L$_4$ are the microscope objectives; 
PH is the pinhole;
NLC is the NLC cell; 
P$_1$ and P$_2$ are the polarizers; 
W$_1$ and W$_2$ are the quarter wave plates;
CCD is the CCD camera.
A microscope objective L$_3$ is illuminated with an elliptically polarized and expanded
parallel beam of light from a He-Ne laser. The output from a second
objective L$_4$ is collected by a CCD camera through the Stokes analyzer.
}
\label{fig:setup}
\end{figure*}

\section{Experiment}
\label{sec:experiment}

In our experiments we used the NLC cells
of thickness $d=110$~\mum 
filled with the nematic liquid crystal mixture E7 from Merk. 
Two glass substrates 
were assembled to form a hometropically oriented NLC cell.
At the wavelength of light generated by a low power He-Ne laser
from Coherent Group
with $\lambda=632.8$~nm (see Fig.~\ref{fig:setup}),
the ordinary and extraordinary refractive indices of 
the NLC are $n_o=1.5246$ and $n_e=1.7608$, respectively;
the refractive index of the glass substrates is $n_g=1.5$.

Figure~\ref{fig:setup} shows our  experimental setup 
devised   to perform the conoscopic measurements using the
Stokes polarimetry 
technique~\cite{Soskin:optl:2003,Soskin:jopa:2004,Dennis:prl:2005,Kis:ujph:2007}.
Referring to Fig.~\ref{fig:setup},
the cell is irradiated with a convergent light beam
formed by the  microscope objective of high numerical aperture~L$_3$. 
The input polarizer P$_1$ is combined with
the properly oriented quarter wave plate W$_1$ to control 
the polarization characteristics (the ellipticity and the azimuth of polarization) 
of a He-Ne laser beam which is expanded and collimated using the
lenses L$_1$ 
and L$_2$. 
A charge coupled device (CCD) camera collects the output from
the microscope objective L$_4$ through the collimating lense L$_5$ 
and the Stokes analyzer represented by the combination
of the quarter wave plate W$_2$ and the polarizer P$_2$. 

This optical arrangement therefore collects simultaneously
the transmittance of the cell for a range of incident angles.
The distribution of the Stokes parameters $S_0\ldots S_3$
describing the state of polarization of the transmitted light
can then be obtained by performing the measurements at
six different combinations of the quarter wave plate W$_2$ and the analyzer P$_2$ 
and using the well-known relations~\cite{Born}
\begin{subequations}
\label{eq:Stokes-exp}
\begin{align}
S_1&=I_{0^\circ}-I_{90^\circ},\\
S_2&=I_{45^\circ}-I_{135^\circ},\\
S_3&=I_\text{RCP}-I_\text{LCP},\\
S_0&=\sqrt{S_1^2+S_2^2+S_3^2},
\end{align}
\end{subequations}
where $I_{0^\circ}$, $I_{90^\circ}$, $I_{45^\circ}$ and $I_{135^\circ}$ 
are the linearly polarized components
(subscript indicates the orientation angle of the analyzer); 
$I_\text{RCP}$ and $I_\text{LCP}$ are the right- and left-handed 
circular polarized components measured
in the presence of the quarter wave plate W$_2$. 
 
Geometrically, the distribution of the Stokes parameters
measured in the observation plane 
can be conveniently represented by the two-dimensional field
of the polarization ellipses.
The geometrical elements of polarization ellipses 
that determine the polarization ellipse field 
can be readily computed from the Stokes parameters~\eqref{eq:Stokes-exp}. 

The orientation of a polarization ellipse is specified 
by the azimuthal angle of polarization
(\textit{polarization azimuth})
\begin{equation}
\phi_\text{p}=\frac{1}{2}\arg(S_1+iS_2)
\label{eq:azimuth}
\end{equation}
and its eccentricity is described by
the signed ellipticity parameter
\begin{equation}
\epsilon_{\ellpt}=\tan\left[\frac{1}{2}\arcsin\left(\frac{S_3}{S_0}\right)\right].
\label{eq:epsilon}
\end{equation}
This parameter will be referred to as the \textit{ellipticity}. 
The handedness of the ellipse is determined by the
sign of the ellipticity parameter $\epsilon_{\ellpt}$. 

In Sec.~\ref{sec:results},
we shall describe
the polarization ellipse fields representing 
the experimentally measured angular patterns.
These will be compared with the predictions of the theory
discussed in the subsequent section.

\section{Theory}
\label{sec:theory}

\subsection{Transmission boundary-value problem}
\label{subsec:transm-problem}

We consider a nematic liquid crystal (NLC) cell of thickness $d$ 
sandwiched between two
parallel plates that are normal to the $z$ axis: $z=0$ and $z=d$.
Typically, anisotropy of nematics is locally uniaxial
and NLC molecules align on average along a local unit 
director,$\uvc{d}$~\cite{Gennes:bk:1993}.
In this case the NLC director
\begin{align}
  \label{eq:director-d}
  \uvc{d}\equiv\vc{e}_{0}(\uvc{d})=\sin\theta_{\dd}\cos\phi_{\dd}\,\uvc{x}
+\sin\theta_{\dd}\sin\phi_{\dd}\,\uvc{y}
+\cos\theta_{\dd}\,\uvc{z}
\end{align}
determines the optic axis.
So, the uniaxially anisotropic dielectric tensor, $\bs{\varepsilon}$, is given by
\begin{align}
  \label{eq:diel-tens-uniax}
  \bs{\varepsilon}=\epsilon_{\perp}\,\mvc{I}_3+\Delta\epsilon\,\uvc{d}\otimes\uvc{d},
\quad
\Delta\epsilon=\epsilon_{\parallel}-\epsilon_{\perp},
\end{align}
where 
$\mvc{I}_n$ is the $n\times n$ identity matrix.
(In what follows carets will denote unit vectors.)
Its two principal values $\epsilon_{\perp}$ and
$\epsilon_{\parallel}$
define the ordinary and extraordinary refractive indices, 
$n_o=\sqrt{\mu\epsilon_{\perp}}$ and
 $n_e=\sqrt{\mu\epsilon_{\parallel}}$,
where $\mu$ is the NLC magnetic permeability.

In a more general case of biaxial 
nematics~\cite{Luckh:thsf:2001,Luckh:nat:2004}
that were recently observed
experimentally~\cite{Madsen:prl:2004,Severing:prl:2004},
there are three different dielectric constants $\epsilon_1$, $\epsilon_2$ and
$\epsilon_3$ representing the eigenvalues of the dielectric
tensor:
\begin{align}
  \label{eq:diel-tens-biax}
  \bs{\varepsilon}=\epsilon_3\,\mvc{I}_3+\Delta\epsilon_1\,\uvc{d}\otimes\uvc{d}
+\Delta\epsilon_2\,\uvc{m}\otimes\uvc{m},
\end{align}
where $\Delta\epsilon_i=\epsilon_i-\epsilon_3$,
and the eigenvectors $\uvc{d}$, $\uvc{m}$ and 
$\uvc{l}=\uvc{d}\times\uvc{m}$
give the corresponding principal axes.

Similar to the director~\eqref{eq:director-d},
the unit vectors $\uvc{m}$ and 
$\uvc{l}$ can be expressed
in terms of Euler angles as follows
\begin{subequations}
  \label{eq:director}
\begin{align}
&
  \label{eq:director-m}
  \uvc{m}=\cos\gamma_{\dd}\,\vc{e}_{1}(\uvc{d})
+\sin\gamma_{\dd}\,\vc{e}_{2}(\uvc{d}),
\\
&
\label{eq:director-l}
  \uvc{l}=-\sin\gamma_{\dd}\,\vc{e}_{1}(\uvc{d})+
\cos\gamma_{\dd}\,\vc{e}_{2}(\uvc{d}),
\end{align}
\end{subequations}
where
$ \vc{e}_{1}(\uvc{d})=(\cos\theta_{\dd}\cos\phi_{\dd},
\cos\theta_{\dd}\sin\phi_{\dd},
-\sin\theta_{\dd})$
and
$ \vc{e}_{2}(\uvc{d})=(-\sin\phi_{\dd}, \cos\phi_{\dd}, 0)$.

\begin{figure*}[!tbh]
\centering
\resizebox{135mm}{!}{\includegraphics*{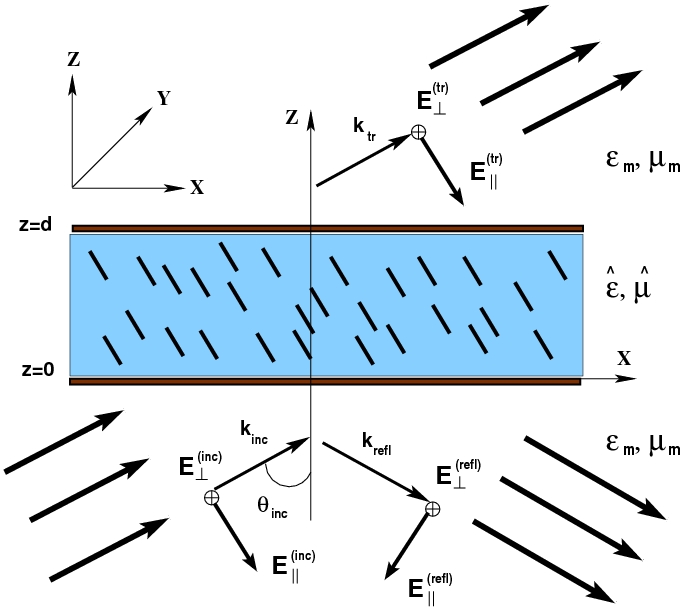}}
\caption{%
Geometry of nematic cell in the plane of incidence.
}
\label{fig:geom}
\end{figure*}

We shall need to write the Maxwell equations for a harmonic
electromagnetic wave (time-dependent factor is $\exp\{-i\omega t\}$)
in the form:
\begin{subequations}
  \label{eq:maxwell}
\begin{align}
&
  \label{eq:maxwell-1}
   \bs{\nabla}\times\vc{E}=i \mu k_{\vac} \vc{H},
\\
&
\label{eq:maxwell-2}
  \bs{\nabla}\times\vc{H}=-i k_{\vac}  \vc{D},
\end{align}
\end{subequations}
where $k_{\vac}=\omega/c$ is the free-space wave number;
$\mu$ is the magnetic permittivity and
$\vc{D}=\bs{\varepsilon}\cdot\vc{E}$ is the electric displacement field.

The medium surrounding the NLC cell is assumed to be optically
isotropic and characterized by the dielectric constant $\epsilon_{\med}$
and the magnetic permittivity $\mu_{\med}$.
So, Maxwell's equations in the medium outside the cell 
can be obtained from Eq.~\eqref{eq:maxwell} by 
replacing $\mu$ and $\vc{D}$ with $\mu_{\med}$ and
$\epsilon_{\med} \vc{E}$, respectively. 

As is shown in Fig.~\ref{fig:geom}, 
there are two plane waves in the half space
$z\le 0$ bounded by the entrance face of the NLC cell: 
the \textit{incoming incident wave} $\{\vc{E}_{\inc}, \vc{H}_{\inc}\}$
and the \textit{outgoing reflected wave} $\{\vc{E}_{\refl}, \vc{H}_{\refl}\}$.
In the half space $z\ge d$ after the exit face, 
the only  wave is the \textit{transmitted plane wave}  $\{\vc{E}_{\trans}, \vc{H}_{\trans}\}$ 
which propagates along the direction of incidence 
and is excited by the incident light. 

So, the electric field outside the cell is 
a superposition of the plane wave solutions
of the Maxwell equations 
\begin{subequations}
  \label{eq:E-med}
\begin{align}
&
  \label{eq:E-before}
  \vc{E}\vert_{z<0}=
\vc{E}_{\inc}(\uvc{k}_{\inc})
\ee^{i(\vc{k}_{\inc}\cdot\vc{r})}
+
\vc{E}_{\refl}(\uvc{k}_{\refl})
\ee^{i(\vc{k}_{\refl}\cdot\vc{r})},
\\
&
 \label{eq:E-after}
  \vc{E}\vert_{z>d}=
\vc{E}_{\trans}(\uvc{k}_{\trans})
\ee^{i(\vc{k}_{\trans}\cdot\vc{r})},
\end{align}
\end{subequations}
where the wave vectors
$\vc{k}_{\inc}$, $\vc{k}_{\refl}$ and $\vc{k}_{\trans}$
that are constrained to lie in the plane of incidence 
due to the boundary conditions requiring 
the tangential components of the electric and magnetic
fields to be continuous at the boundary surfaces.
These conditions are given by 
\begin{subequations}
  \label{eq:bc-gen}
\begin{align}
&
  \label{eq:bc-E}
  \mvc{P}(\uvc{z})\cdot\bigl[
\vc{E}\vert_{z=0+0}-\vc{E}\vert_{z=0-0}
\bigr]
=  \mvc{P}(\uvc{z})\cdot\bigl[
\vc{E}\vert_{z=d+0}-\vc{E}\vert_{z=d-0}
\bigr]=0,
\\
&
 \label{eq:bc-H}
  \mvc{P}(\uvc{z})\cdot\bigl[
\vc{H}\vert_{z=0+0}-\vc{H}\vert_{z=0-0}
\bigr]
=  \mvc{P}(\uvc{z})\cdot\bigl[
\vc{H}\vert_{z=d+0}-\vc{H}\vert_{z=d-0}
\bigr]=0,
\end{align}
\end{subequations}
where
$\mvc{P}(\uvc{z})=\mvc{I}_3-\uvc{z}\otimes\uvc{z}$
is the projector onto 
the plane with the normal directed along the vector $\uvc{z}$
(the $x$-$y$ plane). 

Another consequence of the boundary conditions~\eqref{eq:bc-gen}
is that the tangential components of the wave vectors are the same.
Assuming that the incidence plane is the $x$-$z$
plane we have
\begin{align}
  \label{eq:k-alp}
\vc{k}_{\alpha}=k_{\vac}\vc{q}_{\alpha}=k_{\med}\uvc{k}_{\alpha}=
k_x\,\uvc{x}+
k_z^{(\alpha)}\,\uvc{z},
\quad
\alpha\in\{\inc, \refl, \trans\},   
\end{align}
where $k_{\med}/k_{\vac}=n_{\med}=\sqrt{\mu_{\med}\epsilon_{\med}}$
is the refractive index of the ambient medium
and the components can be expressed
in terms of the incidence angle
$\theta_{\inc}$ as follows
\begin{align}
&
  \label{eq:k_x}
  k_x=k_{\med}\sin\theta_{\inc}\equiv k_{\vac}\, q_x,
\\
&
\label{eq:k_z-alp}
k_z^{(\inc)}=k_z^{(\trans)}=-k_z^{(\refl)}=k_{\med}\cos\theta_{\inc}\equiv
k_{\vac}\,q_z^{(\med)}.
\end{align}
The plane wave 
traveling in the isotropic ambient medium
along the wave vector~\eqref{eq:k-alp} is transverse, so that
the polarization vector is given by 
\begin{align}
&
  \label{eq:E-alp}
  \vc{E}_{\alpha}(\uvc{k}_{\alpha})=
  E_{\parallel}^{(\alpha)}\vc{e}_1(\uvc{k}_{\alpha})+
  E_{\perp}^{(\alpha)}\vc{e}_2(\uvc{k}_{\alpha}),
\\
&
\label{eq:e_x-alp}
\vc{e}_1(\uvc{k}_{\alpha})=
k_{\med}^{-1}
\bigl(
k_z^{(\alpha)}\,\uvc{x}-k_x\,\uvc{z}
\bigr),
\quad
\vc{e}_2(\uvc{k}_{\alpha})=\uvc{y},
\end{align}
where $E_{\parallel}^{(\alpha)}$ and
$E_{\perp}^{(\alpha)}$ are the in-plane and out-of-plane
components of the electric field, respectively.
The vector characterizing the magnetic field is
\begin{align}
\label{eq:H-alp}
  \mu_{\med}\, \vc{H}_{\alpha}(\uvc{k}_{\alpha})=
\vc{q}_{\alpha}\times\vc{E}_{\alpha}(\uvc{k}_{\alpha})=
n_{\med}\,
\bigl[
  E_{\parallel}^{(\alpha)} \uvc{y}-
  E_{\perp}^{(\alpha)} \vc{e}_1(\uvc{k}_{\alpha})
\bigr],
\end{align}
where $\vc{q}_{\alpha}=k_{\vac}^{-1} \vc{k}_{\alpha}=
n_{\med}\uvc{k}_{\alpha}$.
Note that, for plane waves, the dimensionless vector
\begin{equation}
  \label{eq:q-def}
  \vc{q}=k_{\vac}^{-1}\vc{k}
\end{equation}
is parallel to $\vc{k}$ and its length gives the refractive index.
For convenience, we shall often use this vector
in place of the wave vector.

The expressions~\eqref{eq:E-med}--\eqref{eq:H-alp}
give the electromagnetic field of incident,
transmitted and reflected waves propagating in the ambient medium.
This field is of the general form 
\begin{align}
&
  \label{eq:EH-form}
\left\{  
\vc{E}(\vc{r}), \vc{H}(\vc{r})
\right\}
=
\left\{  
\vc{E}(z), \vc{H}(z)
\right\}
\exp[i \sca{\vc{k}_{P}}{\vc{r}}],
\quad
  \vc{k}_{P}=\mvc{P}(\uvc{z})\cdot\vc{k},
\\
&
\label{eq:q_P}
\vc{q}_{P}=k_{\vac}^{-1}\,\vc{k}_{P}=
q_{P}
\left[
\cos(\phi_{\inc})\,\uvc{x}+\sin(\phi_{\inc})\,\uvc{y}
\right],
\end{align}
where the azimuthal angle $\phi_{\inc}$ specifies
orientation of the incidence plane. So, in our case,
we have $\phi_{\inc}=0$ and $q_{P}=q_x$.

The representation~\eqref{eq:EH-form} can be applied to describe 
the field inside the cell when the dielectric
tensor is independent of the in-plane coordinates $x$ and $y$.
In this case the lateral components of the 
electric and magnetic fields
\begin{align}
&
  \label{eq:E_in-plane}
  \vc{E}_{P}=\mvc{P}(\uvc{z})\cdot\vc{E}=
E_{x}\uvc{x}+E_{y}\uvc{y}\equiv
\begin{pmatrix}
  E_x\\
E_y
\end{pmatrix},
\\
&
\label{eq:H_in-plane}
  \vc{H}_{P}=\vc{H}\times\uvc{z}=
H_{y}\uvc{x}-H_{x}\uvc{y}\equiv
\begin{pmatrix}
  H_y\\
-H_x
\end{pmatrix}
\end{align}
can be conveniently combined into the vector
\begin{align}
\label{eq:F-def}
  \vc{F}(z)\equiv \begin{pmatrix}
\vc{E}_{P}(z)\\ \vc{H}_{P}(z)
\end{pmatrix}=
  \begin{pmatrix}
E_x(z)\\ E_y(z)\\H_y(z)\\-H_x(z)
\end{pmatrix}.  
\end{align}
From Maxwell's equations~\eqref{eq:maxwell},
the components normal to the substrates, $E_z$ and $H_z$,
can be readily expressed in terms of the lateral components
\begin{align}
  \label{eq:Ez-E_P}
  \mu H_z=
\sca{\vc{q}_{P}}{\vc{E}_{P}\times\uvc{z}},
\quad
\epsilon_{zz} E_z=
-\sca{\uvc{z}}{\bs{\varepsilon}\cdot\vc{E}_{P}}
-
\sca{\vc{q}_{P}}{\vc{H}_{P}}.
\end{align}
By using the relations~\eqref{eq:Ez-E_P} to eliminate the
normal components we obtain
the equation for the field vector~\eqref{eq:F-def}
in the following matrix form
\begin{align}
  \label{eq:matrix-system}
  -i\pdrs{\tau}\vc{F}=\mvc{M}\cdot\vc{F}\equiv 
    \begin{pmatrix}\mvc{M}_{11}&\mvc{M}_{12}\\\mvc{M}_{21}&\mvc{M}_{22} \end{pmatrix}
    \begin{pmatrix}\vc{E}_{P}\\\vc{H}_{P} \end{pmatrix},
\quad
\tau\equiv k_{\vac} z,
\end{align}
where
\begin{subequations}
\label{eq:matrix-M}
    \begin{align}
&
\label{eq:M-11}
      M^{(11)}_{\alpha\beta}=-\epsilon_{zz}^{-1}
      q^{(P)}_{\alpha} \epsilon_{z\beta},
\\
&
\label{eq:M-22}
      M^{(22)}_{\alpha\beta}=-\epsilon_{zz}^{-1}
      \epsilon_{\alpha z}
      q^{(P)}_{\beta},
\\
&
\label{eq:M-12}
       M^{(12)}_{\alpha\beta}=\mu\delta_{\alpha\beta} 
-\epsilon_{zz}^{-1}q^{(P)}_{\alpha}q^{(P)}_{\beta},
\\
&
\label{eq:M-21}
       M^{(21)}_{\alpha\beta}=\epsilon_{\alpha\beta} 
- \mu^{-1} p^{(P)}_{\alpha}p^{(P)}_{\beta}
-\epsilon_{zz}^{-1}\epsilon_{\alpha z}\epsilon_{z\beta},
    \end{align}
\end{subequations}
where 
$\delta_{\alpha\beta}$ is the Kroneker symbol,
$\vc{q}_{P}=k_{\vac}^{-1}\,\vc{k}_{P}$ and
$\vc{p}_{P}=\uvc{z}\times\vc{q}_{P}$.

The field vector~\eqref{eq:F-def} can now be written 
in the form of a general solution to the linear problem~\eqref{eq:matrix-system}
\begin{align}
  \label{eq:oper-evol}
  \vc{F}(\tau)=\mvc{U}(\tau)\cdot\vc{F}(0),
\end{align}
where $\mvc{U}(\tau)$ is the \textit{evolution operator} 
that can be determined by solving 
the initial value problem
\begin{align}
  \label{eq:eq-oper-evol}
  -i\pdrs{\tau}\mvc{U}=\mvc{M}\cdot\vc{U},
\quad 
\mvc{U}(0)=\mvc{I}_4.
\end{align}

Our above notations can be regarded as a version of 
the well known matrix formalism due to Berreman~\cite{Berreman:josa:1972}.
By using these notations we derive
the boundary conditions~\eqref{eq:bc-gen} 
in the matrix form
\begin{subequations}
  \label{eq:bc-matr}
\begin{align}
&
  \label{eq:bc-matr-in}
  \vc{F}_{<}=\mvc{V}_{\med}
\begin{pmatrix}
\vc{E}_{\inc}\\ \vc{E}_{\refl}
\end{pmatrix}
=\vc{F}(0),
\\
&
\label{eq:bc-matr-out}
\vc{F}_{>}=\mvc{V}_{\med}
\begin{pmatrix}
\vc{E}_{\trans}\\ \vc{0}
\end{pmatrix}
=\vc{F}(h),\quad
h\equiv k_{\vac}\,d,
\end{align}
\end{subequations}
where
$
\vc{E}_{\alpha}\equiv\begin{pmatrix}
E_{\parallel}^{(\alpha)}\\ E_{\perp}^{(\alpha)}
\end{pmatrix}
$.

The matrix $\mvc{V}_{\med}$ 
relates the field vectors,
$\vc{F}_{<}\equiv\vc{F}(0-0)$ and $\vc{F}_{>}\equiv\vc{F}(h+0)$,
and the vector amplitudes $\vc{E}_{\alpha}$
of the waves in the surrounding medium.
From Eqs.~\eqref{eq:E-alp}-\eqref{eq:H-alp},
its block structure 
\begin{align}
  \label{eq:Vm-block}
  \mvc{V}_{\med}=
\begin{pmatrix}
\mvc{E}_{\med} & -\bs{\sigma}_3 \mvc{E}_{\med}\\
\mvc{H}_{\med} & \bs{\sigma}_3 \mvc{H}_{\med}\\
\end{pmatrix}
\end{align}
is characterized by the two diagonal $2\times 2$ matrices
\begin{align}
&
\label{eq:HE_med}  
  \mvc{E}_{\med}=\diag(q_z^{(\med)}/n_{\med}, 1),
\quad
  \mu_{\med}\,\mvc{H}_{\med}=\diag(n_{\med}, q_z^{(\med)}),
\end{align}
where $q_z^{(\med)}=\sqrt{n_{\med}^2-q_x^2}$ and $\bs{\sigma}_3=\diag(1,-1)$.

It is rather starightforward to check the validity of the
algebraic identity for the matrix~\eqref{eq:Vm-block}
\begin{align}
  \label{eq:orth-matr}
  \tcnj{[\mvc{V}_{\med}]}\cdot\mvc{G}\cdot\mvc{V}_{\med}=\mvc{N}_{\med}
=N_{\med}\diag(\mvc{I}_2,-\mvc{I}_2),
\quad
\mvc{G}\equiv\begin{pmatrix}\vc{0} & \mvc{I}_2\\
\mvc{I}_2 & \vc{0}
\end{pmatrix},
\end{align}
where 
$N_{\med}=2q_z^{(\med)}/\mu_{\med}$ 
and
the superscript $T$ indicates matrix transposition.

According to Ref.~\cite{Oldano:pra:1989},
identities of the form~\eqref{eq:orth-matr} can be derived as 
the orthogonality relations
resulted from the conservation law for the energy flux in non-absorbing media.
Algebraically, Eq.~\eqref{eq:orth-matr} can be used to simplify
 inversion of the matrix $\mvc{V}_{\med}$
 \begin{align}
   \label{eq:inv_Vm}
   \mvc{V}_{\med}^{-1}=
\mvc{N}_{\med}^{-1}\cdot\tcnj{[\mvc{V}_{\med}]}\cdot\mvc{G}
 \end{align}
 and to ease qualitative analysis  
[see , e.g., Ref.~\cite{Yakovl:opsp:1998}
for a more extended discussion of applications].

After substituting Eq.~\eqref{eq:oper-evol} into the
boundary conditions~\eqref{eq:bc-matr-in}
we have
\begin{align}
&
  \label{eq:rel-oper}
  \begin{pmatrix}
\vc{E}_{\inc}\\ \vc{E}_{\refl}
\end{pmatrix}
=\mvc{W}
\begin{pmatrix}
\vc{E}_{\trans}\\ \vc{0}
\end{pmatrix},
\\
&
  \label{eq:W-op}
  \mvc{W}=
\mvc{V}_{\med}^{-1}\cdot\mvc{U}^{-1}(h)\cdot\mvc{V}_{\med}=
\begin{pmatrix}
\mvc{W}_{11} & \mvc{W}_{12}\\ 
\mvc{W}_{21} & \mvc{V}_{22}
\end{pmatrix},
\end{align}
where $\mvc{U}^{-1}\equiv \mvc{U}^{-1}(h)$.

From Eq.~\eqref{eq:rel-oper} we have
\begin{align}
  \label{eq:transm-rel}
  \begin{pmatrix}
E_{\parallel}^{(\trans)}\\
E_{\perp}^{(\trans)}
\end{pmatrix}
&
=\mvc{T}
\begin{pmatrix}
E_{\parallel}^{(\inc)}\\
E_{\perp}^{(\inc)}
\end{pmatrix},
\\
  \label{eq:trans-mat1}
 \mvc{T}
&
=\mvc{W}_{11}^{-1},
\end{align}
where $\mvc{T}$ is the \textit{transmission (transfer) matrix}
linking the transmitted and incident waves.

Similar result for the reflected wave is
\begin{align}
&
  \label{eq:refl-rel}
  \begin{pmatrix}
E_{\parallel}^{(\refl)}\\
E_{\perp}^{(\refl)}
\end{pmatrix}
=
\mvc{R}
\begin{pmatrix}
E_{\parallel}^{(\inc)}\\
E_{\perp}^{(\inc)}
\end{pmatrix},
\\
&
  \label{eq:refl-mat1}
 \mvc{R}=\mvc{W}_{21}\cdot\mvc{W}_{11}^{-1}=\mvc{W}_{21}\cdot\mvc{T},
\end{align}
where $\mvc{R}$ is the \textit{reflection matrix}.

For uniformly anisotropic media, the elements of the matrix~\eqref{eq:matrix-M}
are constants and solving Eq.~\eqref{eq:eq-oper-evol}
gives the formula for the evolution matrix
\begin{align}
&
  \label{eq:oper-evolv-uni}
  \mvc{U}(\tau)=\exp\{i \mvc{M} \tau \}
=
\mvc{V}\cdot\exp\{i \bs{\Lambda} \tau \}
\cdot\mvc{V}^{-1},
\\
&
\label{eq:eigv-M-def}
\bs{\Lambda}=\diag(\lambda_1,\lambda_2,\lambda_3,\lambda_4),\quad
\mvc{M}\cdot\mvc{V}=\mvc{V}\cdot\bs{\Lambda},
\end{align}
where $\lambda_i$ is the eigenvalue of the matrix $\mvc{M}$
and $\mvc{V}=(\vc{V}_1\,\vc{V}_2\,\vc{V}_3\,\vc{V}_4)$ 
is the matrix composed of the right eigenvectors,
$\mvc{M}\cdot\vc{V}_i=\lambda_i\,\vc{V}_i$.
Technical details on computing the evolution operator
for both biaxial and uniaxial anisotropies
can be found in Appendix~\ref{sec:evolution_oper}.
In this section we restrict ourselves to 
the simplest (and important for our purposes) case 
that occurs when 
the NLC cell is homeotropically aligned. 

For the homeotropic director structure with $\uvc{d}=\uvc{z}$,
the eigenvalue problem for the matrix $\mvc{M}$
is easy to solve.
The result is
\begin{align}
&
  \label{eq:eigv_homeot}
  \bs{\Lambda}=\diag(q_z^{(e)}, q_z^{(o)},-q_z^{(e)}, -q_z^{(o)}),
\quad
\mvc{V}=
\begin{pmatrix}
\mvc{E}_{+} & -\bs{\sigma}_3 \mvc{E}_{+}\\
\mvc{H}_{+} & \bs{\sigma}_3 \mvc{H}_{+}\\
\end{pmatrix},
\\
&
\label{eq:EpHp-hom}
\mvc{E}_{+}=\diag(q_z^{(e)}/n_{\perp},1),\quad
\mu \mvc{H}_{+}=\diag(n_{\perp},q_z^{(o)}),
\end{align}
where $q_{z}^{(o)}=\sqrt{n_{\perp}^2-q_x^2}$ and
$q_{z}^{(e)}=n_{\perp} n_{\parallel}^{-1}\sqrt{n_{\parallel}^2-q_x^2}$.

The analytical relations~\eqref{eq:eigv_homeot} can now be
substituted  into the expression~\eqref{eq:oper-evolv-uni}
to yield the evolution matrix $\mvc{U}^{-1}(h)=\mvc{U}(-h)$
that enter the linking matrix $\mvc{W}$
defined in Eq.~\eqref{eq:W-op}.
The final step involves substituting the block $2\times 2$ 
matrices $\mvc{W}_{11}$ and $\mvc{W}_{21}$
into Eqs.~\eqref{eq:trans-mat1} and~\eqref{eq:refl-mat1} 
to find the transfer and the transmission matrices
in the following form
\begin{align}
&
  \label{eq:T-homeotr}
  \mvc{T}=\diag(t_e,t_o)=
\Bigl[
\cos(\bs{\Lambda}_{+} h) -  i\bs{\Gamma}_{+}\cdot\sin(\bs{\Lambda}_{+} h)
\Bigr]^{-1},
\\
&
  \label{eq:R-homeotr}
  \mvc{R}=\diag(r_e,r_o)=
-i\bs{\sigma}_{3}\cdot\bs{\Gamma}_{-}\cdot\sin(\bs{\Lambda}_{+} h)\cdot\mvc{T},
\end{align} 
where
\begin{align}
&
  \label{eq:delta-homeotr}
\bs{\Lambda}_{+} h =\diag(q_z^{(e)}, q_z^{(o)}) h\equiv\diag(\delta_e,\delta_o),
\\
&
  \label{eq:gamma-homeotr}
\bs{\Gamma}_{\pm}=(\bs{\Gamma}\pm\bs{\Gamma}^{-1})/2,
\quad
\bs{\Gamma}=
\frac{\mu_{\med}}{\mu}
\begin{pmatrix}
  \dfrac{n_{o}^2 q_z^{(\med)}}{n_{\med}^2 q_z^{(e)}} & 0\\
0 & \dfrac{q_z^{(o)}}{q_z^{(\med)}}
\end{pmatrix}\equiv\diag(\gamma_e,\gamma_o).  
\end{align}

\begin{figure*}[!tbh]
\centering
\subfigure[Star: $I_{C}=-1/2$ and $N_{C}=3$]{%
\resizebox{50mm}{!}{\includegraphics*{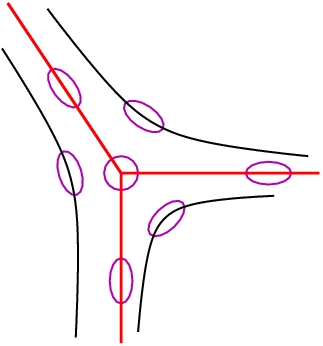}}
\label{fig:star}
}
\subfigure[Lemon: $I_{C}=+1/2$ and $N_{C}=1$]{%
\resizebox{50mm}{!}{\includegraphics*{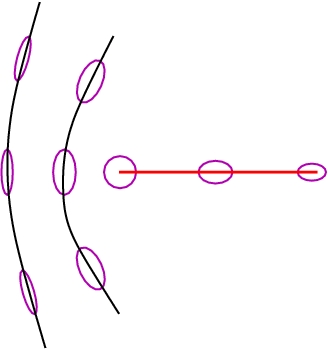}}
\label{fig:lemon}
}
\subfigure[Monstar: $I_{C}=+1/2$ and $N_{C}=3$]{%
\resizebox{50mm}{!}{\includegraphics*{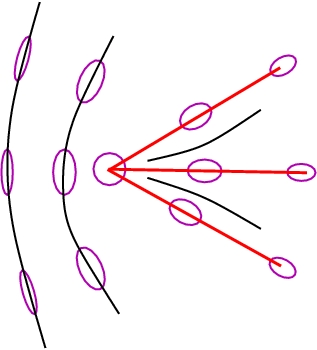}}
\label{fig:monstar}
}
\caption{%
Arrangment of the polarization ellipses around
the C-points of three different types.
}
\label{fig:singul-types}
\end{figure*}

\subsection{Polarization resolved angular patterns}
\label{subsec:pol-resolv-pattern}

We will now proceed  to a study of 
how the polarization properties of the transmitted
wave depend on the direction of the incident beam.
This direction is specified by two angles: the incidence angle $\theta_{\inc}$
and the azimuthal angle of the plane of incidence $\phi_{\inc}$.

In the previous section 
and in Appendix~\ref{sec:evolution_oper},
the transmission problem was analyzed
in the plane of incidence where $\phi_{\inc}=0$.
Clearly, when $\phi_{\inc}\ne 0$,  
in order to have the Euler angles
describing orientation of the director~\eqref{eq:director-d}
with respect to the incidence plane 
we need to replace
the director azimuthal angle $\phi_{\dd}$ with
$\phi_{\dd}-\phi_{\inc}$.

In this section
dependence of the polarization parameters of transmitted waves
on the angles $\theta_{\inc}$ and $\phi_{\inc}$ will be of our primary
concern.
For this purpose, we assume that 
the transmission matrix
$\mvc{T}(\theta_{\inc},\theta_{\dd},\phi_{\dd})$
considered in Sec.~\ref{subsec:transm-problem}
and in Appendix~\ref{sec:evolution_oper}
is changed to
$\mvc{T}(\theta_{\inc},\theta_{\dd},\phi_{\dd}-\phi_{\inc})$.

As previously discussed in Sec.~\ref{sec:experiment},
experimentally, the characteristics of the polarization ellipse
can be determined by measuring 
the Stokes parameters~\eqref{eq:Stokes-exp}.
These parameters are related to 
the \textit{coherence matrix} with the elements
$\mvc{M}_{\alpha \beta}=E_{\alpha} \cnj{E_{\beta}}$,
where $\alpha,\beta\in\{\parallel,\perp\}$~\cite{Born:bk:1999,Azz:1977}.
In the circular basis,
$\sqrt{2}\,\vc{e}_{\pm}(\uvc{k})= \vc{e}_1(\uvc{k})
\pm i \vc{e}_2(\uvc{k})$, 
this matrix written
as a linear combination of the Pauli matrices
gives the Stokes parameters, $S_i$, as its
coefficients
\begin{align}
&
  \label{eq:coher-matr}
  \mvc{M}_c=\mvc{C}\cdot\mvc{M}\cdot\hcnj{\mvc{C}}=
\begin{pmatrix}
|E_{+}|^2 & E_{+} \cnj{E}_{-}\\
E_{-} \cnj{E}_{+} & |E_{-}|^2
\end{pmatrix}
= 2^{-1}\sum_{i=0}^4
S_i\,\bs{\sigma}_i,
\end{align}
where
\begin{align}
&
\label{eq:Pauli-matr}
\bs{\sigma}_0=\mvc{I}_2,
\quad
\bs{\sigma}_1=\begin{pmatrix}0 & 1 \\ 1 & 0\end{pmatrix},
\quad
\bs{\sigma}_2=\begin{pmatrix}0 & -i \\ i & 0\end{pmatrix},
\quad
\bs{\sigma}_3=\begin{pmatrix}1 & 0 \\ 0 & -1\end{pmatrix},
\\
&
  \label{eq:circ-basis}
 \begin{pmatrix}E_{+} \\  E_{-}\end{pmatrix}
=\mvc{C}\cdot 
\begin{pmatrix}E_{\parallel} \\  E_{\perp}\end{pmatrix},
\quad 
\mvc{C}=2^{-1/2}\begin{pmatrix}1 & -i \\ 1 & i\end{pmatrix}. 
\end{align}
 
Since the determinant of the coherence matrix 
vanishes, $\det\mvc{M}=0$,
the Stokes parameters lie on the four-dimensional cone
$S_0^2=\sum_{i=1}^3 S_i^2$,
and can be parameterized as follows
\begin{subequations}
  \label{eq:Stokes}
\begin{align}
&
  \label{eq:S_0}
  S_0 = |E_{+}|^2+|E_{-}|^2=|E_{\parallel}|^2+|E_{\perp}|^2,
\\
&
\label{eq:S_1}
  S_1 = 2\Re\cnj{E}_{+} E_{-}=|E_{\parallel}|^2-|E_{\perp}|^2
=S_0\cos 2\chi_{p}\cos 2\phi_{p},
\\
&
\label{eq:S_2}
  S_2 = 2\Im\cnj{E}_{+} E_{-}=2\Re{E}_{\perp} \cnj{E}_{\parallel}
=S_0\cos 2\chi_{p}\sin 2\phi_{p},
\\
&
 \label{eq:S_3}
  S_3 =|E_{+}|^2-|E_{-}|^2 =2\Im{E}_{\perp} \cnj{E}_{\parallel}
=S_0\sin 2\chi_{p},
\end{align}
\end{subequations}
where $0<\phi_p\le\pi$ is the polarization 
azimuth~\eqref{eq:azimuth}
and $-\pi/4 \le\chi_p\le \pi/4$ is the ellipticity angle.
Then, the relations expressing the ellipse characteristics
in terms of the Stokes parameters are 
\begin{align}
&
  \label{eq:ell-az-S12}
  \phi_p=2^{-1} \arg(\cnj{E}_{+} E_{-})=
2^{-1}\arg{S},\quad S\equiv S_1+ i S_2,
\\
&
\label{eq:ellip-param-S3}
\epsilon_{\ellpt}=
\frac{|E_{+}|-|E_{-}|}{|E_{-}|+|E_{+}|}
=
\tan\chi_p,\quad
\chi_p = 2^{-1}\arcsin(S_3/S_0).
\end{align}

Similar to Refs.~\cite{Dennis:optcom:2002,Jacks:bk:1999},
we have used Eq.~\eqref{eq:coher-matr} and Eq.~\eqref{eq:S_3}
to define the Stokes parameter $S_3$ which is opposite in sign
to that given in the book~\cite{Born}. 
The ellipse is considered to be
right-handed (RH) if 
its helicity is positive, so that $|E_{+}|>|E_{-}|$ and $\epsilon_{\ellpt}>0$.
In the opposite case with $\epsilon_{\ellpt}<0$,
the ellipse is left-handed (LH).

In the spherical basis, $\{\vc{e}_{+}, \vc{e}_{-}\}$,
similar to Eq.~\eqref{eq:transm-rel}, 
the transmission matrix
\begin{align}
  \label{eq:T-circul}
  \mvc{T}_c=
\begin{pmatrix}
t_{++} & t_{+-}\\ t_{-+} & t _{--}
\end{pmatrix}
={\mvc{C}}\cdot\mvc{T}\cdot\hcnj{\mvc{C}}
\end{align}
relates the circular components of the incident and transmitted waves
\begin{align}
  \label{eq:transm_circ-rel}
  \begin{pmatrix}
E_{+}^{(\trans)}\\ E_{-}^{(\trans)}
\end{pmatrix}
=
  \mvc{T}_c\cdot
  \begin{pmatrix}
E_{+}^{(\inc)}\\ E_{-}^{(\inc)}
\end{pmatrix},
\end{align}
where
$E_{\pm}^{(\alpha)}=(E_{\parallel}^{(\alpha)}\mp i E_{\perp}^{(\alpha)})/\sqrt{2}$. 
 So, for the transfer matrix
describing the conoscopic patterns
on the transverse plane of projection,
we have~\cite{Kis:arxiv:2006,Kis:jpcm:2007}
\begin{align}
&
  \label{eq:tilde-T}
  \tilde{\mvc{T}}(\rho,\phi)
=\exp(-i\phi\,\bs{\sigma}_3)\cdot\mvc{T}_c(\rho,\phi)\cdot\exp(i\phi\,\bs{\sigma}_3),
\\
&
\label{eq:rho}
\rho= r \tan\theta_{\inc},\quad
\phi= \phi_{\inc},
\end{align} 
where $\rho$ and $\phi$ are the polar coordinates
in the observation plane
($x=\rho\cos\phi$ and $y=\rho\sin\phi$
are the Cartesian coordinates)
and 
$r$ is the aperture dependent scale factor.

For the elliptically polarized incident plane wave
with the circular components
\begin{align}
  \label{eq:circ_inc_waves}
 E_{\nu}^{(\inc)}=\exp\bigl(- i \nu\phi_{p}^{(\inc)}\bigr) 
\bigl[1+\nu\epsilon_{\ellpt}^{(\inc)}\bigr] 
|E_{\inc}|, 
\end{align}
expressed in terms of the polarization azimuth
$\phi_{p}^{(\inc)}$
and the ellipticity parameter $\epsilon_{\ellpt}^{(\inc)}$, 
the reduced components of the transmitted wave are given by
\begin{align}
&
  \label{eq:E_mu_tr}
  E_{\nu}^{(\trans)}/|E_{\inc}|\equiv\Psi_{\nu}=
|\Psi_{\nu}| \ee^{i\phi_{\nu}}=
\Bigl\{
t_{\nu,\,\nu} 
\bigl[
1+\nu\epsilon_{\ellpt}^{(\inc)}
\bigr]
+ 
\notag
\\
&
t_{\nu,\,-\nu}
\bigl[
1-\nu\epsilon_{\ellpt}^{(\inc)}
\bigr]
\exp(-2i\nu\psi)
\Bigr\}
\exp(-i\nu\phi_p^{(\inc)}),
\end{align}
where $\psi=\phi-\phi_p^{(\inc)}$.

For the transmitted wave, 
the formula~\eqref{eq:E_mu_tr} gives
the normalized Stokes parameters
$s_i^{(\trans)}=S_i^{(\trans)}/S_0^{(\trans)}$
as a function of the incidence angles, $\theta_{\inc}$
and $\phi_{\inc}$.
From Eqs.~\eqref{eq:ell-az-S12} and~\eqref{eq:ellip-param-S3},
these parameters are determined by the ellipticity,
$\epsilon_{\ellpt}=(|\Psi_{+}|-|\Psi_{-}|)/(|\Psi_{+}|+|\Psi_{-}|)$,
and the polarization azimuth, $\phi_p=(\phi_{-}-\phi_{+})/2$.
Owing to the relation~\eqref{eq:rho},
the incidence angles and the points in the observation plane
are in one-to-one correspondence.
So,
the distribution of the normalized Stokes parameters
can be evaluated by 
computing the characteristics of the polarization ellipse 
$\epsilon_{\ellpt}$ and $\phi_p$ at 
each point of the projection plane.
Geometrically, this procedure 
yields the two-dimensional field of polarization ellipses
that might be called the
\textit{polarization resolved angular (conoscopic) pattern}.
 
The point where
$|\Psi_{\nu}|=0$ and thus
the transmitted wave is circularly polarized
with $\epsilon_{\ellpt}=-\nu$ 
will be referred to as  the \textit{C$_{\nu}$-point}.
This is an example of 
the \textit{polarization singularity} 
where the phases $\phi_{\nu}$ and $\phi_{p}$
become indeterminate.

Equivalently, C-points can be viewed as 
the \textit{phase singularities} of the complex scalar field
\begin{align}
  \label{eq:tld-S}
  \tilde{S}=\cnj{\Psi}_{+}\Psi_{-}=\tilde{S}_1+i\tilde{S}_2
\end{align}
proportional to the Stokes field defined in Eq.~\eqref{eq:ell-az-S12}.
Such singularities are characterized by 
the \textit{winding number} which is the signed number
of rotations  of the two-component field
$(\tilde{S}_1,\tilde{S}_2)$ around the circuit surrounding the
singularity~\cite{Merm:rpm:1979}.
The winding number also known as the \textit{signed strength of the
dislocation} is generically $\pm 1$.

Since the polarization azimuth~\eqref{eq:ell-az-S12} is defined modulo
$\pi$ and $2\phi_p=\arg\tilde{S}$, the dislocation strength is twice
the index of the corresponding C$_{\nu}$-point, $I_{C}$.
For generic C points, $I_{C}=\pm 1/2$
and the index can be computed from the formula
\begin{align}
  \label{eq:I_c-gen}
  I_{C}=\frac{1}{2}\sign
\Bigr[
\Im(\pdrs{x}\cnj{\tilde{S}}\pdrs{y}\tilde{S})
\Bigr]_{\substack{x=x_{\nu}\\ y=y_{\nu}}},
\end{align}
where $\pdrs{x} f$ is the partial derivative of $f$ with respect to $x$.

The relation~\eqref{eq:I_c-gen} gives the index of the C$_{\nu}$-point with
the coordinates $(x_{\nu},y_{\nu})$
expressed in terms of the vorticity~\cite{Dennis:prsl:2000,Dennis:optcom:2002}:
$\Im(\pdrs{x}\cnj{\tilde{S}}\pdrs{y}\tilde{S})
=\pdrs{x}\tilde{S}_1\pdrs{y}\tilde{S}_2-
\pdrs{y}\tilde{S}_1\pdrs{x}\tilde{S}_2
$.
The formula linking gradients of
the complex field 
$\sca{\vc{E}}{\vc{E}}\propto\Psi_{+}\Psi_{-}$
and the index for C-lines in the three-dimensional space
was derived in Ref.~\cite{Berry:jopa2:2004}.

In the case with $\Psi_{\nu}=0$ (C$_{\nu}$-point), 
only derivatives of $\Psi_{\nu}$ enter the
expression~\eqref{eq:I_c-gen}
which can be suitably rearranged to yield 
the index of the C$_{\nu}$-point in the following form:
\begin{align}
  \label{eq:I_c-Psi_mu}
  I_{C}=\frac{\nu}{2}\sign
\Bigr[
\Im(\pdrs{x}\Psi_{\nu}\,\pdrs{y}\cnj{\Psi}_{\nu})
\Bigr]_{\substack{x=x_{\nu}\\ y=y_{\nu}}}
=
\frac{\nu}{2}\sign
\Bigr[
\Im(\pdrs{\rho}\Psi_{\nu}\,\pdrs{\phi}\cnj{\Psi}_{\nu})
\Bigr]_{\substack{\rho=\rho_{\nu}\\ \phi=\phi_{\nu}}}.
\end{align}
Subsequently, we shall apply
the formula~\eqref{eq:I_c-Psi_mu} expressing the index in terms of
the derivatives with respect to polar coordinates 
to the case of homeotropically aligned cell.

In addition to the handedness and the index,
the C-points are classified according to the number of 
straight lines terminating on the singularity.
This is the so-called \textit{line classification}
that was initially studied in the context of 
umbilic points~\cite{Berry:jpa:1977}.

For generic C-points,
as is shown in Fig.~\ref{fig:singul-types},
the number of the straight lines, $N_{C}$, may either be 1 or 3.
This number is $3$ provided the index equals $-1/2$, $I_{C}=-1/2$,
and such C-points are called \textit{stars}.
At $I_{C}=1/2$, there are two characteristic patterns of polarization ellipses
around a C point: (a)~\textit{lemon} with $N_{C}=1$ and 
(b)~\textit{monstar} with $N_{C}=3$~\cite{Nye:prsl:1983a}.
Different quantitative criteria to distinguish between the C-points of
the lemon and the monstar types were deduced
in Refs.~\cite{Dennis:optcom:2002,Kis:jpcm:2007}.

\begin{figure*}[!tbh]
\centering
   \resizebox{165mm}{!}{\includegraphics*{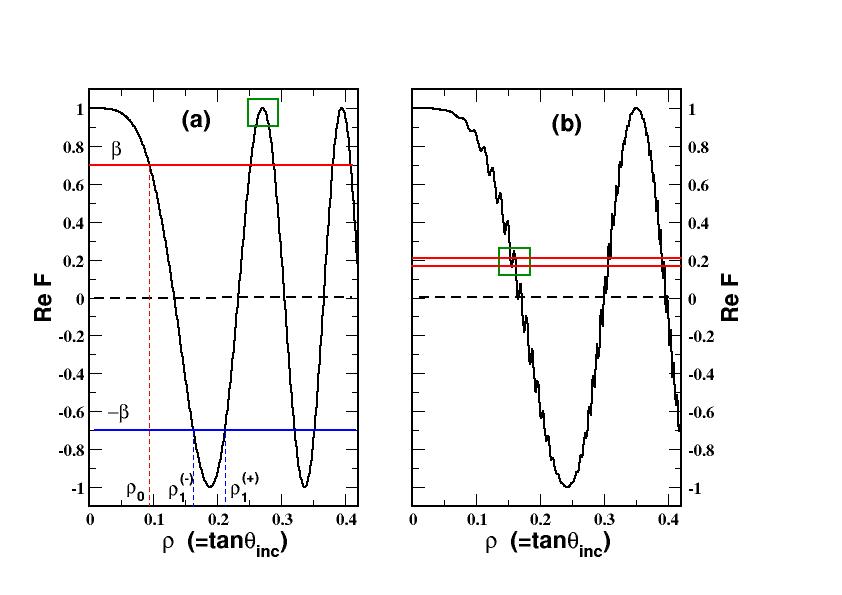}}
 \caption{
The real part of the function~\eqref{eq:F_vs_rho}
that enter the left-hand side of Eq.~\eqref{eq:C-mu-cond}
computed for 
(a)~the cell filled with the NLC mixture E7
and (b)~the KDP (potassium dihydrogen phosphate) birefringent crystal.
The parameters are:
(a)~$n_{\perp}=1.5246$;
$n_{\parallel}=1.7608$;
$n_{\med}=1.5$; 
$d=110$~\mum; and
(b)~$n_{\perp}=1.5093$;
$n_{\parallel}=1.4683$;
$n_{\med}=1.0$;
$d=300$~\mum.
 }
\label{fig:im_f_rho}
\end{figure*}

The location of the C$_{\nu}$-points on the projection plane
is determined by the polar coordinates:
$\rho_{k}^{(\nu)}$ and $\phi_{k}^{(\nu)}$,
where $k$ is the numbering label.
These can be found by solving the equation
\begin{align}
  \label{eq:C-mu-gen}
  |\Psi_{\nu}(\rho,\phi)|=0
\end{align}
that generally has multiple solutions.

The case of linearly polarized wave with $\epsilon_{\ellpt}=0$
provides another example of the polarization singularity
where the handedness is undefined.
The curves along which the polarization is linear are
called the \textit{L-lines}.

The transmitted wave is linearly polarized
when the condition
\begin{align}
  \label{eq:L-lines}
  |\Psi_{+}(\rho,\phi)|=|\Psi_{-}(\rho,\phi)|
\end{align}
is satisfied.
So, Eq.~\eqref{eq:L-lines} describes loci of points forming
the L-lines lying in the projection plane.

\begin{figure*}[!tbh]
\centering
\subfigure[]{
   \resizebox{70mm}{!}{\includegraphics*{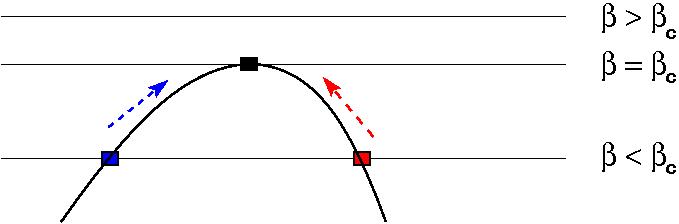}}
\label{subfig:bifurc_1}
}
\subfigure[]{
   \resizebox{70mm}{!}{\includegraphics*{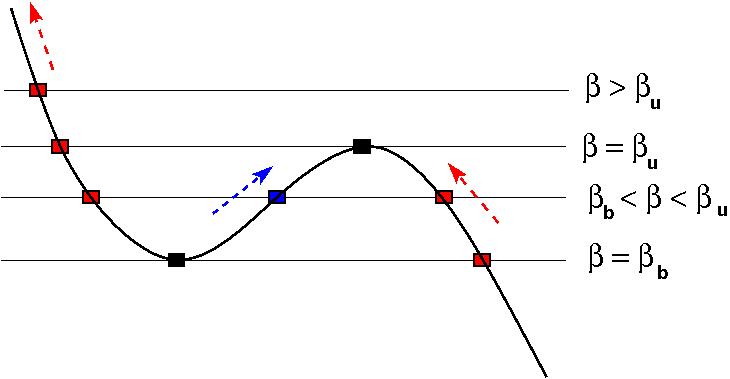}}
\label{subfig:bifurc_2}
}
 \caption{
Schematic respresentation of
creation and annihilation of C-points
governed by saddle-node bifurcations of the solutions
of Eq.~\eqref{eq:C-mu-cond}
that take place within the areas 
marked by the boxes in 
Figs.~\ref{fig:im_f_rho}(a) and~\ref{fig:im_f_rho}(b).
(a)~C-points annihilate as the governing parameter 
$\beta$ passes through its critical value
$\beta_c$ in the immediate vicinity of unity.
(b)~Two successive bifurcations
result in
creation and annihilation of C-points at 
$\beta=\beta_b$ and $\beta=\beta_u$, respectively.
 }
\label{fig:bifurcation}
\end{figure*}

\begin{figure*}[!tbh]
\centering
\subfigure[$\epsilon_{\ellpt}^{(\inc)}=0.0$]{
   \resizebox{65mm}{!}{\includegraphics*{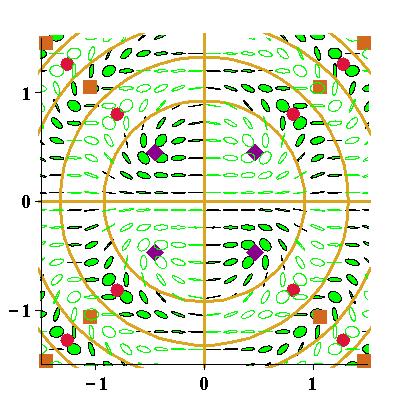}}
\label{subfig:th_00}
}
\subfigure[$\epsilon_{\ellpt}^{(\inc)}=0.01$]{
   \resizebox{65mm}{!}{\includegraphics*{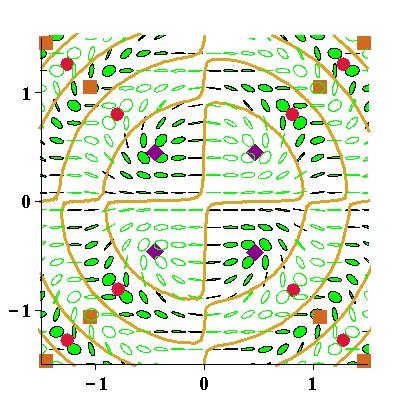}}
\label{subfig:th_01}
}
\subfigure[$\epsilon_{\ellpt}^{(\inc)}=-0.01$]{
   \resizebox{65mm}{!}{\includegraphics*{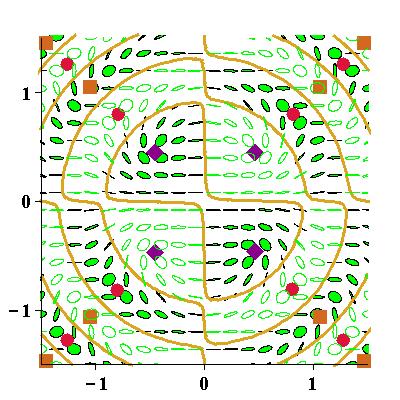}}
\label{subfig:th_01m}
}
 \caption{Polarization resolved conoscopic patterns 
computed as polarisation ellipse fields 
in the observation plane for 
the homeotropically oriented cell filled with the NLC mixture E7
at small values of the ellipticity:
(a)~$\epsilon_{\ellpt}^{(\inc)}=0.0$,
(b)~$\epsilon_{\ellpt}^{(\inc)}=0.01$,
and (c)~$\epsilon_{\ellpt}^{(\inc)}=-0.01$.
The parameters used in calculations are:
  the wavelength is $\lambda_{\inc}=632.8$~nm; 
the polarization azimuth is $\phi_p^{(\inc)}=0$;
 the cell thickness is $d=110\text{ }\mu$m;
the scale factor is $r=5$~mm;
$n_{\med}=1.5$, $n_{\perp}=1.5246$, and $n_{\parallel}=1.7608$.  
The C-points of the star, the lemon and the monstar
 types are marked by circles, diamonds and squares, respectively. 
L-lines are represented by solid lines.  Left-handed and right-handed
polarization is respectively indicated by solid and open ellipses.
 }
\label{fig:avoid_cross}
\end{figure*}

\begin{figure*}[!tbh]
\centering
\subfigure[$\epsilon_{\ellpt}^{(\inc)}=\epsilon_2^{(c)}-10^{-7}$]{
   \resizebox{65mm}{!}{\includegraphics*{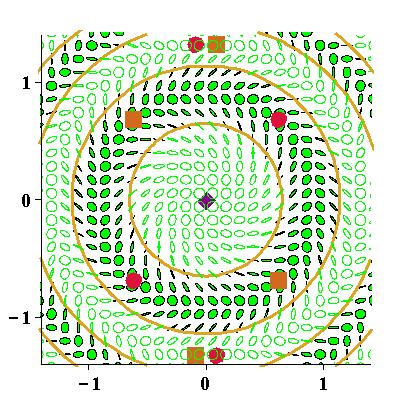}}
\label{subfig:under_e2}
}
\subfigure[$\epsilon_{\ellpt}^{(\inc)}=\epsilon_1^{(c)}-10^{-7}$]{
   \resizebox{65mm}{!}{\includegraphics*{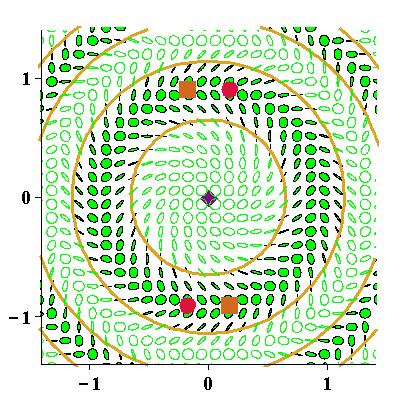}}
\label{subfig:under_e1}
}
\subfigure[$\epsilon_{\ellpt}^{(\inc)}-=\epsilon_1^{(c)}+10^{-6}$]{
   \resizebox{65mm}{!}{\includegraphics*{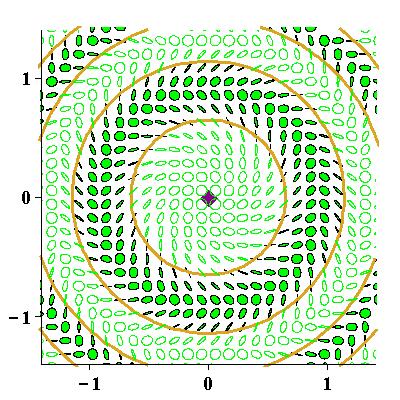}}
\label{subfig:above_e1}
}
 \caption{Polarization resolved conoscopic patterns 
of the homeotropically oriented NLC cell  
computed at different values of the ellipticity.
Three cases are shown:
(a)
$\epsilon_3^{(c)}<\epsilon_{\ellpt}^{(\inc)}<\epsilon_2^{(c)}$;
(b)
$\epsilon_2^{(c)}<\epsilon_{\ellpt}^{(\inc)}<\epsilon_1^{(c)}$;
(c)
$\epsilon_1^{(c)}<\epsilon_{\ellpt}^{(\inc)}<1$.
The list of other parameters is given in the caption of Fig.~\ref{fig:avoid_cross}.
 }
\label{fig:annihilation}
\end{figure*}

\subsection{Ellipticity induced effects}
\label{subsec:ellipt-dynam}

\subsubsection{C-points: creation and annihilation}
\label{subsubsec:c-points}

When the director is normal to the substrates,
the NLC cell is homeotropically aligned and $d_z=1$.
In this case the transmission matrix~\eqref{eq:T-homeotr} 
is diagonal and its elements in the circular basis
are given by
\begin{align}
  \label{eq:t_pm}
  t_{\nu,\nu}=t_{+},\quad
  t_{\nu,-\nu}=t_{-},\quad
t_{\pm}=
(t_{e}\pm t_{o})/2, 
\end{align}
where the transmission coefficients 
\begin{align}
&
  \label{eq:t_eo}
  t_{e,\,o}=t_{e,\,o}(q_x)=
\bigl[
\cos(\delta_{e,\,o})- i \gamma_{e,\,o}\,\sin(\delta_{e,\,o})
\bigr]^{-1}
\equiv t_{e,\,o}(\rho)
\end{align}
are expressed in terms of the phases
\begin{align}
  \label{eq:delta_eo}
  \delta_{e}=
q_{z}^{(e)} h
=
n_{\perp} n_{\parallel}^{-1}\sqrt{n_{\parallel}^2-q_x^2}
\,h,
\quad
\delta_{o}=
q_{z}^{(o)} h
=\sqrt{n_{\perp}^2-q_x^2}\, h
\end{align}
and the amplitudes $\gamma_{e,\,o}$
defined in Eq.~\eqref{eq:gamma-homeotr}.

Owing to the cylindrical symmetry of the homeotropic structure,
the transmission coefficients~\eqref{eq:t_eo}
do not depend on the azimuthal angle of the incidence plane
$\phi\equiv\phi_{\inc}$.
From Eq.~\eqref{eq:E_mu_tr}
another consequence of this symmetry
is that the azimuth, $\phi_{p}-\phi_{p}^{(\inc)}$,
and the ellipticity of the transmitted wave
depend only on 
the difference of the azimuthal angles: $\phi-\phi_{p}^{(\inc)}$.
It follows that 
the sole effect of changing the polarization azimuth of the incident wave
is the rotation of the polarization ellipse field by the angle $\phi_{p}^{(\inc)}$.
So,  we focus our attention on the effects 
governed by the ellipticity of the incident wave, $\epsilon_{\ellpt}^{(\inc)}$.

From Eqs.~\eqref{eq:E_mu_tr} and~\eqref{eq:t_pm},
the C$_{\nu}$-points may appear only if
the condition
\begin{align}
  \label{eq:C_nu_cond1}
\frac{|t_{\nu,-\nu}|}{|t_{\nu,\nu}|}\equiv
\frac{|t_{-}|}{|t_{+}|}=
\frac{1+\nu\epsilon_{\ellpt}^{(\inc)}}{1-\nu\epsilon_{\ellpt}^{(\inc)}}
 \end{align}
is satisfied.
Another form of this condition is
\begin{align}
&
  \label{eq:C-mu-cond}
  \Re{F(\rho)}=-\nu\,\frac{2\epsilon_{\ellpt}^{(\inc)}}{1+[\epsilon_{\ellpt}^{(\inc)}]^2}\equiv
\epsilon_{\ellpt}\, \mu_{\inc}\, \beta,
\\
&
\label{eq:F_vs_rho}
F(\rho)=
\frac{2\, t_e(\rho)\,\cnj{t}_o(\rho)}{|t_e(\rho)|^2+|t_o(\rho)|^2}=
\frac{\exp(i\Delta)}{\cosh(\ln |t_e|-\ln |t_o|)},
\end{align}
where
$\beta=2|\epsilon_{\ellpt}^{(\inc)}|(1+[\epsilon_{\ellpt}^{(\inc)}]^2)^{-1}$,
$\Delta=\arg(t_e\cnj{t}_o)$,
$\epsilon_{\ellpt}=-\nu$ is the ellipticity of the C$_{\nu}$-point
and $\mu_{\inc}=\sign(\epsilon_{\ellpt}^{(\inc)})$ is the 
handedness (helicity) of the incident wave.
Solutions of equation~\eqref{eq:C-mu-cond}
define the radii of circles containing the C-points.

Plots depicted in Fig.~\ref{fig:im_f_rho}
demonstrate non-monotonic
behaviour of the function 
$\Re F$ that enter
the left-hand side of Eq.~\eqref{eq:C-mu-cond}.

The experimentally relevant 
case of the nematic cell filled with the LC mixture E7
is shown in Fig.~\ref{fig:im_f_rho}(a).
It is characterized by the weak anisotropy
of the transmission amplitudes
$|t_e|$ and $|t_o|$,
so that $|1-|t_e|/|t_o||< 10^{-4}$
for the incidence angles up to $60$ degrees.
For such cells, the phase of the function~\eqref{eq:F_vs_rho},
$\Delta$, can be approximated 
by  the difference in optical path
of the ordinary and extraordinary waves,
$\delta=\delta_e-\delta_o$.
So, the formula 
\begin{align}
  \label{eq:F_approx}
  F(\rho) 
\approx F_0(\rho)=\exp(i \delta),
\quad
\delta=\delta_e-\delta_o,
\end{align}
provides a high accuracy approximation
for the expression~\eqref{eq:F_vs_rho}
and, for the LC mixture E7, the deviation
$\max|F-F_0|$ can be estimated to be better than
$5.0\times 10^{-4}$.
 
The radii of the circles with the C-points.  
can now be accurately located by solving 
the approximate equation $\cos \delta=\pm \beta$
obtained from Eq.~\eqref{eq:C-mu-cond}
by using the expression~\eqref{eq:F_approx}.
The result can written as two sequences of radii
\begin{align}
  \label{eq:rho_pmk}
|\delta(\rho_0)|\approx \arccos(\beta),\quad
  |\delta(\rho_k^{(\pm)})|\approx \pm\arccos(\beta)+\pi k,\quad
k=1,\ldots, N,
\end{align}
where 
$N$ is determined by the number of solutions and
the circles of each sequence, $\rho_k^{(+)}$ and $\rho_k^{(-)}$, 
are numbered
by the non-negative integer $k$
in non-decreasing order of size, so that
$\rho_0\equiv\rho_{0}^{(\pm)}\le\rho_{1}^{(-)}\le\rho_{1}^{(+)}\ldots\le\rho_{k}^{(-)}\le\rho_{k}^{(+)}\le\ldots$.

At $\rho=\rho_k^{(\pm)}$, 
there is a pair of the C-points 
on each circle of the radius $\rho_k^{(\pm)}$
with the ellipticity and the azimuthal angles given by
\begin{align}
&
  \label{eq:epsl_k}
  \epsilon_{\ellpt}\Bigr|_{\rho=\rho_k^{(\alpha)}}
=(-1)^k \mu_{\inc}\equiv \mu_k,
\quad
k=0,\ldots, N,
\\
&
  \label{eq:phi-mu}
  \phi_{\pm k}^{(\alpha)}=\phi_p^{(\inc)}\pm\pi/2-
\frac{\mu_k}{2}
\arg\Bigl[\cnj{t}_{+} t_{-}\Bigr]_{\rho=\rho_k^{(\alpha)}},
\end{align}
where $\rho_{0}^{(\alpha)}\equiv \rho_{0}$ and $\alpha \in\{+,-\}$.

The symmetric arrangment of the C-points 
is a consequence of the symmetry relation
\begin{align}
  \label{eq:symmetry-homeo}
  |\Psi_{\nu}(\rho,\psi)|=|\Psi_{\nu}(\rho,\pi+\psi)|
\end{align}
for the amplitudes~\eqref{eq:E_mu_tr}.

We can now substitute Eq.~\eqref{eq:E_mu_tr} into the
expression for the index~\eqref{eq:I_c-Psi_mu}
and use the relations~\eqref{eq:t_pm} and~\eqref{eq:C-mu-cond}
to recast the result into the form of the index formula
\begin{align}
  \label{eq:I_c-homeo}
  I_{C}=-\frac{1}{2}\,\sign \Bigl[\pdrs{\rho}
  \Re{F(\rho)}\Bigr]_{\rho=\rho_k^{(\pm)}}=\frac{\pm (-1)^{k}}{2}\,,
\quad
k=0, \ldots N,
\end{align}
where the second equality follows because
$\Re F$ is an oscillating function of $\rho$ and $F(0)=1$.

Interestingly, Eq.~\eqref{eq:I_c-homeo} relates the index of C-points
and derivatives of the transmission coefficients with
respect to the incidence angle 
($\pdrs{\rho} f =\cos^2\theta_{\inc}\,\partial
f/\partial\theta_{\inc}$).
The index is determined by the circle number $k$
and alternates in sign starting from $I_{C}=+1/2$
which is the index of two 
C-points symetrically arranged in the vicinity of the origin.

So, the pair of C-points lying on the smallest circle of the radius 
$\rho_0$ can either be of lemon or monstar types.
Interestingly, when 
$\beta$ approaches unity and the incident light becomes
circularly polarized, the radius $\rho_0$ vanishes and the C-points
merge into one C-point of the index $+1$.

This is the limiting case where the incident light is circular
polarized with $\epsilon_{\ellpt}^{(\inc)}=\pm 1$
and a C-point may develop provided that
the transmission coefficients meet 
the isotropy condition: $t_e = \pm t_o$. 
As it can be seen
from Eq.~\eqref{eq:t_eo} we have $t_e = t_o$ for the case of normal
incidence with $q_x=\sin\theta_{\inc}=0$. 
The corresponding C-point is located at the origin
and, as it was discussed earlier, its index $I_{C}$ equals $+1$. 

At this stage, however, 
it remains unclear if the above-mentioned C-point is unique. 
In order to clarify this,
we shall look more closely at the events
that happen when the ellipticity parameter $\beta$
defined in Eq.~\eqref{eq:C-mu-cond}
varies from zero to unity.
At $\beta=0$, the incident light 
is in the state of linear polarization
with $\epsilon_{\ellpt}^{(\inc)}=0$, whereas
the incident wave is circular polarized with
$|\epsilon_{\ellpt}^{(\inc)}|=1$ at $\beta=1$.  

The case of linearly polarized incident waves
was previously studied in Refs.~\cite{Kis:arxiv:2006,Kis:jpcm:2007}.
Fig.~\ref{subfig:th_00} shows the polarization resolved
angular pattern as the polarization ellipse field
computed at $\beta=0$.
It can be seen that we have a sequence of $N+1$ concentric circles
each containing two pairs of symmetrically arranged
C-points.
The index and the handedness of the C-points
alternate in sign along the radial direction.
The C-points on the circle of the smallest
radius, $\rho=\rho_0$, corresponding to the direction close 
to the normal incidence 
are found to be of the lemon type with the index equal to $+1/2$.

Algebraically,
from Eq.~\eqref{eq:rho_pmk}
it follows that
the difference between the radii of the C-points 
$\rho_{k}^{(+)}$ and $\rho_{k+1}^{(-)}$,
where $k=0,\ldots,N$, vanishes with
the ellipticity $\epsilon_{\ellpt}^{(\inc)}$ 
approaching zero.
So, at $\epsilon_{\ellpt}^{(\inc)}=0$, the angular polarization structure 
is characterized by the presence of degeneracy
which is a consequence of
the additional symmetry relation
\begin{align}
  \label{eq:symmetry-homeo-lin}
 |\Psi_{\nu}(\rho,\psi)|=|\Psi_{-\nu}(\rho,\pi-\psi)|.
\end{align}

The symmetry-breaking effects due to non-vanishing ellipticity
of the incident wave
are illustrated in Figs.~\ref{subfig:th_01} and~\ref{subfig:th_01m}
for small values of $\epsilon_{\ellpt}^{(\inc)}$.
This is the case when the relation~\eqref{eq:symmetry-homeo-lin}
breaks down removing the above degeneracy, 
so that $\rho_{k}^{(+)}<\rho_{k+1}^{(-)}$
at $\epsilon_{\ellpt}^{(\inc)}\ne 0$.

As is schematically illustrated by Fig.~\ref{subfig:bifurc_1}, 
there is another type of degeneracy which takes place
when the ellipticity parameter $\beta$
reaches a local maximum of the function $\Re F$
such as one marked by box in Fig.~\ref{fig:im_f_rho}(a).
In contrast to the above-discussed case of linear polarization,
the number of C-points changes 
as the parameter $\beta$ and the ellipticity $\epsilon_{\ellpt}^{(\inc)}$
pass through their critical values, $\beta_c$ and $\epsilon^{(c)}$.

The picture shown in Fig.~\ref{subfig:bifurc_1} is typical
of a saddle-node (tangent) bifuraction~\cite{Hohen:rmp:1977,Devaney:bk:1989}.
Such bifurcations occur at the incidence angles 
represented by 
the values of the radii (see Eq.~\eqref{eq:rho})
at local extrema of the real part of the function~\eqref{eq:F_vs_rho}.
By applying the approximation~\eqref{eq:F_approx},
these radii can be estimated from the solutions of the equation
\begin{align}
  \label{eq:rho_crit}
  |\delta(\rho_k^{(c)})|\approx\pi k,\quad
k=0,\ldots,N
\end{align}
obtained from Eq.~\eqref{eq:rho_pmk} 
by setting $\arccos(\beta)=0$.

Given the critical value of the radius, $\rho=\rho_k^{(c)}$,
we can use Eq.~\eqref{eq:C-mu-cond} 
to deduce the relations 
\begin{align}
  \label{eq:beta_c}
  \beta_k^{(c)}=
\bigl|
\Re{F(\rho_k^{(c)})}
\bigr|,
\quad
\beta_k^{(c)} \epsilon_k^{(c)}=
-1+\sqrt{
1-  \left[\beta_k^{(c)}\right]^2
}
\end{align}
and to compute the corresponding values of the
parameter  $\beta$ and the ellipticity:
$\beta=\beta_k^{(c)}$ and 
$|\epsilon_{\ellpt}^{(\inc)}|=\epsilon_k^{(c)}$.
It is also not difficult to show that $\rho_0^{(c)}=0$ and
$\rho_k^{(-)}\le \rho_k^{(c)}\le \rho_k^{(+)}$.

Our analysis suggests that the ellipse field undergoes
a sequence of tangent bifurcations 
as the magnitude of 
the ellipticity $|\epsilon_{\ellpt}^{(\inc)}|$
increases from zero to unity.
There are two pairs of the C-points
with $\rho=\rho_{k}^{(-)}$ and $\rho=\rho_{k}^{(+)}$
that merge and, subsequently, annihilate
as the governing parameter
passes through the bifurcation point
with $|\epsilon_{\ellpt}^{(\inc)}|=\epsilon_k^{(c)}$
and $\rho_{k}^{(+)}=\rho_{k}^{(-)}$.

In addition,
an important consequence of the index formula~\eqref{eq:I_c-homeo}
is that the indicies of the C-points involved in a tangent bifurcation  
differ only in sign,  
so that the total sum of indicies remains intact. 
Obviously, this result agrees 
with the conservation law of the total topological index.

Note, however, that, 
within the approximation~\eqref{eq:F_approx},
the C-points cannot merge and annihilate
provided the incident wave is not circular polarized.
Mathematically, 
the difficulty is that, for the approximate function 
$\cos\delta$,  the critical values of the ellipticity
are all equal to unity, $\beta_{k}^{(c)}=\epsilon_{k}^{(c)}=1$.
So, we arrive at the conclusion that
the difference $1-\epsilon_{k}^{(c)}$
is determined by the accuracy of 
the approximation~\eqref{eq:F_approx}.  

For the cell filled with the LC mixture E7,
the fisrt three critical values of the ellipticity
$\epsilon_3^{(c)} < \epsilon_2^{(c)}< \epsilon_1^{(c)}$
turned out to be  very close to unity:
$1-\epsilon_1^{(c)}\approx 1.22\times 10^{-6}$,
$1-\epsilon_2^{(c)}\approx 1.17\times 10^{-5}$
and
$1-\epsilon_3^{(c)}\approx 4.16\times 10^{-5}$.
The polarization resolved angular patterns
presented in Fig.~\ref{fig:annihilation}
are computed at the three different values of 
$\epsilon_{\ellpt}^{(\inc)}$:
(a)
$\epsilon_3^{(c)}<\epsilon_{\ellpt}^{(\inc)}=\epsilon_2^{(c)}-10^{-7}<\epsilon_2^{(c)}$;
(b)
$\epsilon_2^{(c)}<\epsilon_{\ellpt}^{(\inc)}=\epsilon_1^{(c)}-10^{-7}<\epsilon_1^{(c)}$;
(c)
$\epsilon_1^{(c)}<\epsilon_{\ellpt}^{(\inc)}=\epsilon_1^{(c)}+10^{-6}<1$.
The pairs of the C-points with $k=3$,
$k=2$ and $k=1$ are shown to coalesce and disappear in succession
as the ellipticity passes sequentially through $\epsilon_3^{(c)}$,
$\epsilon_2^{(c)}$ and $\epsilon_1^{(c)}$.

So, in the limit of circular polarization with
$\epsilon_{\ellpt}^{(\inc)}=1$,
we have the pattern where the only C-point is located at the origin. 
The index of this C-point is $I_{C}= +1$.
This result is in agreement with the Euler-Poincar\'e theorem
linking the Euler characteristics, $\chi$,
of a two-dimesional manifold 
and the total topological charge (index) of a 
smooth vector field on the manifold~\cite{Monast:bk:1993}.
In our case, it can be concluded that
$\chi=2 I_{C}$ is the total index,
where $\chi=+2$ is the Euler characteristics of a sphere.

So far, we considered bifurcations at the local extrema
that, according to Eq.~\eqref{eq:rho_crit}, 
are in one-to-one correspondence with the extrema
of the function $\Re{F_{0}}=\cos\delta$.
Thus, such birfucations are governed by the phase
difference between ordinary and extraordinary waves,
$\delta=\delta_e-\delta_o$.
As is shown in Fig.~\ref{fig:annihilation},
they are typically characterized by 
extremely small differences between
unity and the critical values of the ellipticity.
For this reason they cannot be detected experimentally.

From the above discussion
it can be inferred that
the critical values might be lowered
only if the approximation~\eqref{eq:F_approx} 
does not work well.
In particular, this is the case for
the graph of $\Re{F}$ plotted
in Fig.~\ref{fig:im_f_rho}(b)
where the anisotropic material
is represented by
the KDP (potassium dihydrogen phosphate) birefringent crystal.
The deviation
$\max|F-F_0|$ can be estimated at about
$7.0\times 10^{-2}$, so that $\max\bigl[1-\epsilon_k^{(c)}\bigr]\approx 0.02$.
 
The curve shown in Fig.~\ref{fig:im_f_rho}(b)
also demonstrates that
the breakdown of the approximation~\eqref{eq:F_approx} 
may manifest itself in additional extrema 
formed away from the neighborhood of unity.
As is illustrated in Fig.~\ref{subfig:bifurc_2}
for the region marked by the box in Fig.~\ref{fig:im_f_rho}(b),
these local extrema result in
two successive tangent bifurcations
leading to creation and annihilation of two different
pairs of the C-points.

In contrast to the extrema described by
the phase difference $\delta$,
this is the multireflection which is primarily responsible
for additional small-scale oscillations.
Typically, it is a challenging task 
to resolve accurately these ripple-like and noisy oscillations
in polarimetry measurements.

\subsubsection{Avoided L-line crossings}
\label{subsubsec:avoided-cross}

Now we pass on to discussing the L-lines 
which are the curves of linear polarization
in the observation plane.
Analytically,
substituting Eq.~\eqref{eq:E_mu_tr}
into the relation~\eqref{eq:L-lines}
and using the matrix elements~\eqref{eq:t_pm}
give the equation
\begin{align}
  \label{eq:L-lines-homeo}
2\epsilon_{\ellpt}^{(\inc)}
\Re{F(\rho)}+
(1-[\epsilon_{\ellpt}^{(\inc)}]^2)
  \sin(2\psi)\, \Im{F(\rho)}=0
\end{align}
describing L-lines 
for the case of the homeotropically aligned NLC cell. 

We begin with the results for the case of linearly polarized
incident waves studied in Refs.~\cite{Kis:arxiv:2006,Kis:jpcm:2007}.
At $\epsilon_{\ellpt}^{(\inc)}=0$, from Eq.~\eqref{eq:L-lines-homeo} 
there are two straight lines of linear polarization: $\phi=\phi_p^{(\inc)}$ and
$\phi=\phi_p^{(\inc)}+\pi/2$, where 
the polarization vectors of incident and transmitted waves are parallel,
$\phi_p^{(\trans)}\equiv\phi_p=\phi_p^{(\inc)}$. 
Other L-lines are circles separating the circles of C-points. 
Interestingly,
the radii of the circles, $\rho_{k}^{(L)}$, can be found as the solutions
of Eq.~\eqref{eq:rho_crit} with $k=1,\ldots N$.

When $k$ is even,  it can be concluded that,
similar to the straight L-lines, $\phi_p\approx \phi_p^{(\inc)}$.
If $k$ is odd,  the polarization vector of the transmitted wave 
rotates with the azimuthal
angle of the incidence plane and $\phi_p\approx \phi_p^{(\inc)}+2\phi$.

The structure of the L-lines 
 at $\beta=0$ is shown in Fig.~\ref{subfig:th_00} 
where the L-lines are represented by the solid lines
bounding the regions of left- and right-handed
polarization. 
As it was discussed in the preceding subsection,
the case of linear polarization possesses additional symmetry
described by the relation~\eqref{eq:symmetry-homeo-lin}.
This symmetry is broken at non-zero values of the ellipticity
and the case of linear polarization is structurally unstable.

In Fig.~\ref{subfig:th_01} and Fig.~\ref{subfig:th_01m},
we show that even small values of the ellipticity,
$\epsilon_{\ellpt}^{(\inc)}$,
change the structure of the intersecting L-lines
depicted in Fig.~\ref{subfig:th_00}
into a family of non-intersecting closed L-lines.
The figures clearly demonstrate
what might be called, 
by analogy with the well-known avoided level crossings 
in the quantum mechanics,
the effect of \textit{avoided L-line crossings}

The closed L-lines  gradually evolve into the circles 
as the ellipticity approaches
the limit of circular polarization with
$|\epsilon_{\ellpt}^{(\inc)}|=1$.
The radii of the C-circles $\rho_{k}^{(L)}$
can be found by solving the equation:
$\Re{F}(\rho)=0$.
It follows that 
$\rho_{k}^{(-)}$ at $\epsilon_{\ellpt}^{(\inc)}=0$
is equal to
$\rho_{k}^{(L)}$ at $|\epsilon_{\ellpt}^{(\inc)}|=1$.
So,
we have a family of the concentric circles describing both
the C-points for linearly polarized incident waves
(see Fig.~\ref{subfig:th_00})
and
the L-lines for  circular polarized incident waves.
(see Fig.~\ref{fig:annihilation}). 

\begin{figure*}[!tbh]
\centering
\subfigure[Experiment at $\epsilon_{\ellpt}^{(\inc)}=0.0$]{
   \resizebox{65mm}{!}{\includegraphics*{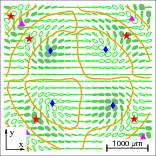}}
}
\subfigure[Theory at $\epsilon_{\ellpt}^{(\inc)}=0.0$]{
   \resizebox{65mm}{!}{\includegraphics*{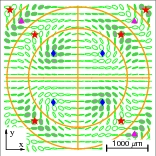}}
}
 \caption{
(a)~Experimentally measured and (b)~theoretically computed
fields of polarization ellipses in the observation plane for the
   homeotropically aligned cell of the thickness $d=110$~\mum\, 
filled with the NLC mixture E7. 
The incident wave is linearly polarized with
the polarization azimuth $\phi_p^{(\inc)}=0$.
The C-points of the
   star, the lemon and the monstar types are marked by stars, diamonds
   and triangles, respectively. L-lines are represented by solid
   lines.  Left-handed and right-handed polarization is respectively
   indicated by solid and open ellipses.}
\label{fig:expth-eps0_0}
\end{figure*}

\begin{figure*}[!tbh]
\centering
\subfigure[Experiment at $\epsilon_{\ellpt}^{(\inc)}=0.26$]{
   \resizebox{65mm}{!}{\includegraphics*{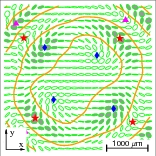}}
}
\subfigure[Theory at $\epsilon_{\ellpt}^{(\inc)}=0.26$]{
   \resizebox{65mm}{!}{\includegraphics*{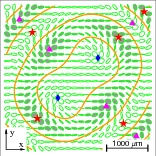}}
}
 \caption{%
(a)~Experimentally measured and (b)~theoretically computed
fields of polarization ellipses for the
 homeotropically aligned cell filled with the NLC mixture E7. 
The ellipticity of the incident light is
$\epsilon_{\ellpt}^{(\inc)}=0.26$.  
}
\label{fig:expth-eps0_26}
\end{figure*}

\begin{figure*}[!tbh]
\centering
\subfigure[Experiment at $\epsilon_{\ellpt}^{(\inc)}=0.65$]{
   \resizebox{65mm}{!}{\includegraphics*{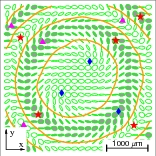}}
}
\subfigure[Theory at $\epsilon_{\ellpt}^{(\inc)}=0.65$]{
   \resizebox{65mm}{!}{\includegraphics*{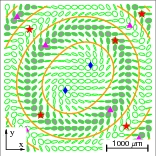}}
}
 \caption{%
(a)~Experimentally measured and (b)~theoretically computed
fields of polarization ellipses for the homeotropically aligned cell 
filled with the NLC mixture E7. 
The ellipticity of the incident wave is 
$\epsilon_{\ellpt}^{(\inc)}=0.65$.
}
\label{fig:expth-eps0_65}
\end{figure*}

\section{Results and discussion}
\label{sec:results}

The analytical results presented in
the previous section completely characterize 
the polarization resolved angular patterns of light transmitted
through the NLC cell 
for the homeotropic orientational structure
in terms of the polarization singularities.
These patterns 
can be conveniently represented as the fields of polarization ellipses
and
characterize the polarization structure behind 
the conoscopic images measured in experiments with two crossed
polarizers. 

After computing loci of the C-points and the L-lines, 
we have deduced a simple formula for the index of 
the C-points~\eqref{eq:I_c-homeo}
and studied what happen to the polarization singularities
if the ellipticity and the polarization azimuth of
the incident wave vary.
It turned out that the C-points may appear and disappear
when the pattern undergoes bifurcations
at some critical values of the ellipticity.
In addition, the patterns formed in the neighborhood of zero
ellipticity (learly polarized incident light)
are characterized by the appearance of avoided L-line crossings.

Now we compare the 
experimentally measured and  the theoretically 
calculated polarization resolved angular 
patterns for the homeotropically oriented NLC cell.
These patterns are shown 
in Figs.~\ref{fig:expth-eps0_0}--\ref{fig:expth-eps0_65}
as the fields of polarization ellipses in 
the observation plane 
where the polar coordinates are defined in
Eq.~\eqref{eq:rho} and the scale factor $r$ is taken to be $5$~mm. 
The parameters used in our calculations are
described in Sec.~\ref{sec:experiment}.
They are also listed in the caption of Fig.~\ref{fig:avoid_cross}.

In
Fig.~\ref{fig:expth-eps0_0},
we present the results obtained for 
the linearly polarized wave
with $\epsilon_{\ellpt}^{(\inc)}=0.0$,
whereas
the cases of elliptically polarized incident light 
with $\epsilon_{\ellpt}^{(\inc)}=0.26$
and $\epsilon_{\ellpt}^{(\inc)}=0.65$
are shown in 
Fig.~\ref{fig:expth-eps0_26}
and
Fig.~\ref{fig:expth-eps0_65},
respectively.
As is evident
from the figures,
the C-points are  arranged in chains formed by four rays 
along which they alternate in sign of the handedness and of the index. 
The L-lines are typically represented by 
the closed curves separating the regions of different polarization handedness.  
It is also clear that the predictions of the theory 
discussed in Sec.~\ref{subsec:ellipt-dynam}
are in good agreement with the experimental data.
 
In addition,
by making a comparison between
the patterns with different values of ellipticity
we arrive at the conclusion that,
when the ellipticity changes, 
the C-points mainly move along the radial 
direction and their azimuthal angles remain
approximately constant. 
Theoretically, it can be shown that,
at $\phi_p^{(\inc)}=0$,
the azimuthal angle $|\phi_{\pm k}^{(\alpha)}|$
given by Eq.~\eqref{eq:phi-mu}
does not depart appreciably
from $\pi/4$ provided the magnitude of 
the ellipticity $|\epsilon_{\ellpt}^{(\inc)}|$
is not in the immediate vicinity of unity.

By contrast,
as is shown in Fig.~\ref{fig:annihilation}, 
annihilation of the C-points is accompanied by
drastic changes in the azimuth of the C-points,
whereas the radius is very close to 
its critical value defined in Eq.~\eqref{eq:rho_crit}. 

The geometric rearrangements
and transformations 
studied in this paper might be regarded as
pseudo-dynamics of the polarization ellipse fields
with the ellipticity served as a ``time'' parameter.
In three-dimensional space, phase singularity lines,
whose geometry depends on  a ``time'' parameter were 
recently investigated in Ref.~\cite{Dennis:jpa:2007}

In conclusion, we note that
the homeotropic alignment  presents 
the simplest case of 
anisotropic orientational structure.
Our theoretical analysis 
can be extended to more complicated cases
involving both uniaxial and biaxial anisotropies
by using the analytical results presented in 
Appendix~\ref{sec:evolution_oper}.
These results will be published elsewhere.

\begin{acknowledgments}
This work was partially supported by CERG Grant No.~612406
and by STCU Grant No.~4687.
 \end{acknowledgments}

\appendix

\section{Operator of evolution}
\label{sec:evolution_oper}

The equations~\eqref{eq:trans-mat1} and~\eqref{eq:refl-mat1}
express the transmission 
and reflection matrices in terms of the linking matrix~\eqref{eq:W-op}
which is determined by  the evolution operator~\eqref{eq:oper-evol}.
In the case of uniform anisotropy, 
the operator can be computed as the evolution
matrix~\eqref{eq:oper-evolv-uni} expressed in terms of
the eigenvalues and the eigenvectors~\eqref{eq:eigv-M-def}
of the matrix~\eqref{eq:matrix-M}.

In this section we consider how
the eigenvalue problem can be solved 
using the basis of \textit{eigenmodes (normal modes)} 
in a uniformly anisotropic medium.
These eigenmodes are known to be
linearly polarized plane waves characterized by 
the $\uvc{k}$ dependent 
refractive indices~\cite{Born,Yariv:bk:1984,Azz:1977}. 

\subsection{Orthogonality relations}
\label{subsec:orth-relat}

From Eqs.~\eqref{eq:M-11}--~\eqref{eq:M-21}, if the dielectric tensor $\bs{\varepsilon}$
is symmetric, $\epsilon_{ij}=\epsilon_{ji}$,
we deduce the relations for the block matrices
\begin{align}
  \label{eq:Mij-symmetr}
  \tcnj{\mvc{M}}_{12}=\mvc{M}_{12},\quad
  \tcnj{\mvc{M}}_{21}=\mvc{M}_{21},\quad
  \tcnj{\mvc{M}}_{11}=\mvc{M}_{22}
\end{align}
giving the symmetry identity 
\begin{align}
  \label{eq:M-symmetry}
  \tcnj{\mvc{M}}=\mvc{G}\cdot\mvc{M}\cdot\mvc{G},
\quad
\mvc{G}=
\begin{pmatrix}
  \mvc{0} & \mvc{I}_2\\
\mvc{I}_2 & \mvc{0}
\end{pmatrix}
\end{align}
for the matrix $\mvc{M}$.

The left eigenvectors, $\{\vc{V}'_1,\vc{V}'_2,\vc{V}'_3,\vc{V}'_4\}$, 
of the matrix $\mvc{M}$
can be defined as the right eigenvectors of its transpose
$\tcnj{\mvc{M}}$
\begin{align}
  \label{eq:left_vect-def}
  \tcnj{\mvc{M}}\cdot\vc{V}'_i=\lambda_i\vc{V}'_i
\Leftrightarrow
\tcnj{\mvc{M}}\cdot\mvc{V}'=\mvc{V}'\cdot\bs{\Lambda},
\end{align}
where $\mvc{V}'=(\vc{V}'_1\,\vc{V}'_2\,\vc{V}'_3\,\vc{V}'_4)$ is 
the matrix of the left eigenvectors.
It is not difficult to prove that the left and right eigenvector
form a biorthogonal set
\begin{align}
  \label{eq:biorthog}
  \sca{\vc{V}'_i}{\vc{V}_j}=
\tcnj{[\vc{V}'_i]}\cdot\vc{V}_j=\delta_{ij} N_i
\Leftrightarrow
\tcnj{\mvc{V}'}\cdot\mvc{V}=\mvc{N},
\end{align}
where $\mvc{N}
=\diag(N_1,N_2,N_3,N_4)$.

Eq.~\eqref{eq:M-symmetry} and the
definition~\eqref{eq:left_vect-def}
can now be combined to yield the relation 
\begin{align}
  \label{eq:left-right-rel}
  \mvc{V}'=\mvc{G}\cdot\mvc{V}
\end{align}
linking the matrices of the left and right eigenvectors,
$\mvc{V}'$ and $\mvc{V}$.
It remains to notice that the identities~\eqref{eq:biorthog}
and~\eqref{eq:left-right-rel} give both the orthogonality
relation
\begin{align}
  \label{eq:orth_rel}
  \tcnj{\mvc{V}}\cdot\mvc{G}\cdot\mvc{V}=\mvc{N}
\equiv\diag(N_1,N_2,N_3,N_4)
\end{align}
and the formula for inverse of the eigenvector matrix
\begin{align}
  \label{eq:inv_V}
  \mvc{V}^{-1}=\mvc{N}^{-1}\cdot\tcnj{\mvc{V}}\cdot\mvc{G}.
\end{align}
The relations given in Eq.~\eqref{eq:orth-matr}
and Eq.~\eqref{eq:inv_Vm}
represent the special case of isotropic media
and, thus, immediately follow from
Eqs.~\eqref{eq:orth_rel} and~\eqref{eq:inv_V}.

By combining the identities~\eqref{eq:inv_Vm} and~\eqref{eq:inv_V}
with the expression for the evolution
operator~\eqref{eq:oper-evolv-uni},
we derive
the linking matrix~\eqref{eq:W-op} for the case of uniform anisotropy
in the following form:
\begin{align}
&
  \label{eq:W-uni}
  \mvc{W}=\mvc{V}_{\med}^{-1}\cdot\exp\{- i \mvc{M} h\}\cdot\mvc{V}_{\med}=
\mvc{N}_{\med}^{-1}\cdot\tilde{\mvc{W}}=
N_{\med}^{-1}
\begin{pmatrix}
  \tilde{\mvc{W}}_{11} & \tilde{\mvc{W}}_{12}\\
  -\tilde{\mvc{W}}_{21} & -\tilde{\mvc{W}}_{22}
\end{pmatrix}
,
\\
&
\label{eq:tildeW}
\phantom{\mvc{W}=\mvc{V}_{\med}^{-1}\cdot}
\tilde{\mvc{W}}=\tcnj{\tilde{\mvc{V}}}\cdot\mvc{W}_d\cdot\tilde{\mvc{V}},
\\
&
\label{eq:tildeV}
\tilde{\mvc{V}}=\tcnj{\mvc{V}}\cdot\mvc{G}\cdot\mvc{V}_{\med},\quad
\mvc{W}_d=\mvc{N}^{-1}\cdot\exp\{-i \bs{\Lambda} h\}.
\end{align}
From Eq.~\eqref{eq:W-uni} and Eq.~\eqref{eq:tildeW} 
the matrices $\tilde{\mvc{W}}$ and $\mvc{W}_{ii}$
are symmetric, whereas $\tcnj{\mvc{W}}_{21}=-\mvc{W}_{12}$. 
So, we conclude that the transmission matrix
$\mvc{T}=\mvc{W}_{11}^{-1}$ is symmetric too.

Note that, for a non-absorbing medium characterized by 
a real matrix $\mvc{M}$, $\Im \mvc{M}=\vc{0}$,
inverse of the matrix~\eqref{eq:W-uni}
can be derived by complex conjugation:
$\mvc{W}^{-1}=\cnj{\mvc{W}}$,
where an asterisk indicates complex conjugation.
In particular,  the conservation law
\begin{align}
  \label{eq:consv_law}
  \hcnj{\mvc{T}}\cdot\mvc{T}+
\hcnj{\mvc{R}}\cdot\mvc{R}=\mvc{I}_2,
\end{align}
where the superscript $\dagger$ stands for Hermitian conjugation,
can be immediately deduced from the relation
\begin{align}
  \label{eq:consv-energy-rel}
  \cnj{\mvc{W}}_{11}\cdot\mvc{W}_{11}+
\cnj{\mvc{W}}_{12}\cdot\mvc{W}_{21}=
  \hcnj{\mvc{W}}_{11}\cdot\mvc{W}_{11}-
  \hcnj{\mvc{W}}_{21}\cdot\mvc{W}_{21}=\mvc{I}_2.
\end{align}

\subsection{Uniformly biaxial media }
\label{subsec:unif-biax-media}

For plane waves, owing to the Maxwell equation~\eqref{eq:maxwell-2},
the electric displacement field is transverse,
$\sca{\vc{D}}{\vc{k}}=0$. So, assuming that the wave
vector $\vc{k}$ lies in the plane of incidence (the $x$-$z$ plane), 
the vector $\vc{D}$ can be conveniently defined by its components
in the basis $\{e_{1}(\uvc{k}), e_{2}(\uvc{k}), e_{0}(\uvc{k})\equiv\uvc{k}\}$
\begin{align}
&
  \label{eq:D-transv}
  \vc{D}=D_{1}\, e_{1}(\uvc{k}) + D_{2}\, e_{2}(\uvc{k}),
\\
&
\label{eq:kxz}
\uvc{k}\equiv e_{0}(\uvc{k})=
k^{-1} \vc{k}=q^{-1} (q_{x}\uvc{x}+q_{z}\uvc{z}),
\end{align}
where 
$e_{1}(\uvc{k})=q^{-1} (q_{z}\uvc{x}-q_{x}\uvc{z})$,
$e_{2}(\uvc{k})=\uvc{y}$ and 
$q=k/k_{\vac}=n$ is the refractive index.

Given the electric displacement field~\eqref{eq:D-transv},
the electric field can be found from the constitutive relation
\begin{align}
&
  \label{eq:D-E}
  \vc{E}=\mu\,\bs{\eta}\cdot\vc{D},
\\
&
\label{eq:tens-eta}
  \bs{\eta}=\eta_3\,\mvc{I}_3+\Delta\eta_1\,\uvc{d}\otimes\uvc{d}
+\Delta\eta_2\,\uvc{m}\otimes\uvc{m},
\end{align}
where $\eta_i=(\mu\epsilon_i)^{-1}$,
$\Delta\eta_i=\eta_i-\eta_3$ and $\mu\bs{\eta}$ is the inverse
dielectric tensor ($\mu\,\bs{\eta}\cdot\bs{\varepsilon}=\mvc{I}_3$).

The Maxwell equations can now be combined with the
relation~\eqref{eq:D-E}
to yield the equation for the displacement vector $\vc{D}$
in the form of an eigenvalue problem 
\begin{align}
  \label{eq:eig-eq}
  \bs{\eta}_{t}\cdot\vc{D}=q^{-2}\vc{D},
\quad
  \bs{\eta}_{t}=\mvc{P}(\uvc{k})\cdot
  \bs{\eta}\cdot\mvc{P}(\uvc{k}),
\end{align}
where
$\mvc{P}(\uvc{k})=\mvc{I}_3-\uvc{k}\otimes\uvc{k}$
is the projector onto 
the plane normal to the wave vector $\vc{k}$.
By using the expression for the inverse dielectric
tensor~\eqref{eq:tens-eta}
we can rewrite the equation~\eqref{eq:eig-eq}
in the explicit matrix form:
\begin{align}
&
  \label{eq:matr-eigeq}
  \bigl[
a_{-}\,\bs{\sigma}_3+b\,\bs{\sigma}_1-c\, \mvc{I}_2
\bigr]
\begin{pmatrix}
  D_{1}\\D_{2}
\end{pmatrix}
=\vc{0},
\quad
\bs{\sigma}_1=
\begin{pmatrix}
  0 & 1\\1 &0
\end{pmatrix},
\\
&
\label{eq:coef-a}
2 a_{\mp} = \Delta\eta_1(\tilde{d}_1^2\mp \tilde{d}_2^2)+
\Delta\eta_2(\tilde{m}_1^2\mp \tilde{m}_2^2),
\\
&
\label{eq:coef-b}
b = \Delta\eta_1\tilde{d}_1 \tilde{d}_2
+\Delta\eta_2\tilde{m}_1 \tilde{m}_2,
\\
&
\label{eq:coef-c}
c=1-\eta_3 q^2-a_{+},
\end{align}
where $\tilde{d}_{i}=q\sca{\uvc{d}}{e_{i}(\uvc{k})}$
 and $\tilde{m}_{i}=q\sca{\uvc{m}}{e_{i}(\uvc{k})}$.
Then the dispersion relation (Fresnel's equation)
\begin{align}
&
   \label{eq:eigval-qz}
\Bigl[
1-\eta_3 q^2
\Bigr]   
\Bigl[
1-\eta_3 q^2-\Delta\eta_1\,
\bigl(
q^2-\sca{\uvc{d}}{\vc{q}}^2
\bigr)-
\notag
\\
&
\Delta\eta_2\,
\bigl(
q^2-\sca{\uvc{m}}{\vc{q}}^2
\bigr) 
\Bigr]-
\Delta\eta_1\,\Delta\eta_2\,q^2
\sca{\uvc{d}\times\uvc{m}}{\vc{q}}^2=0
 \end{align}
can be derived as the condition for 
the system of linear equations~\eqref{eq:matr-eigeq}
to have a non-vanishing solution.

The Fresnel equation describes the wave surface.
In our case solving the algebraic equation~\eqref{eq:eigval-qz}
at $q_x=n_{\med}\sin\theta_{\inc}$
gives the values of the $z$ component of the vector $\vc{q}$, $q_z$.

Generally, there are four roots of Eq.~\eqref{eq:eigval-qz}
$\{q_z^{(1)},q_z^{(2)},q_z^{(3)},q_z^{(4)}\}$
that form a set of the eigenvalues
$\{\lambda_1,\lambda_2,\lambda_3,\lambda_4\}$
defined in Eq.~\eqref{eq:eigv-M-def}.
Each root $q_z^{(i)}$ corresponds to the eigenwave propagating 
inside the cell with the dimensionless wave vector $\vc{q}_{i}=
(q_x,0,q_z^{(i)})$ and the refractive index
$n_{i}=q_{i}$.
The corresponding polarization vector of the electric displacement
field is given by
\begin{align}
  \label{eq:D_i}
  \vc{D}^{(i)}=
\begin{cases}
\cos\phi_{i}\, e_{1}(\uvc{k}_{i}) + 
\sin\phi_{i}\, e_{2}(\uvc{k}_{i}),
& c\vert_{q_z=q_z^{(i)}}>0,
\\
-\sin\phi_{i}\, e_{1}(\uvc{k}_{i}) + 
\cos\phi_{i}\, e_{2}(\uvc{k}_{i}),
& c\vert_{q_z=q_z^{(i)}}<0,
\end{cases}
\end{align}
where
\begin{align}
&
  \label{eq:phi_i}
  2\phi_{i}=\arg(a_{-}+i b)\vert_{q_z=q_z^{(i)}}=
\notag
\\
&
\arg\Bigl[
\Delta\eta_1
(\tilde{d}_1+i \tilde{d}_2)^2
+
\Delta\eta_2
(\tilde{m}_1+i \tilde{m}_2)^2
\Bigr]_{q_z=q_z^{(i)}}.
\end{align}
From Eq.~\eqref{eq:phi_i} it is clear that the azimuthal angle
$\phi_{i}$ becomes indeterminate in the degenerate case
when the coefficients $a_{-}$ and $b$ are both identically 
equal to zero. Typically, as far as the eigenmodes are concerned, 
this case does not present any fundamental difficulties.
It just means that the azimuthal angles of the degenerate eigenmodes
can be prescribed arbitrarily. Such freedom of choice, however, 
does not affect the evolution operator which remains uniquely defined.

The procedure to determine the characteristics
of the eigenmodes involves the following steps:
(a)~evaluation of the eigenvalues $q_z^{(i)}$ by solving the Fresnel
equation~\eqref{eq:eigval-qz};
(b)~calculation of the polarization vectors of the electric
displacement field $\vc{D}^{(i)}$ by 
using the formula~\eqref{eq:D_i};
(c)~computing the polarization vectors of the electric and magnetic fields
from the relations:
$\vc{E}^{(i)}=\mu\,\bs{\eta}\cdot\vc{D}^{(i)}$
(see Eq.~\eqref{eq:D-E}) and 
$\mu\,\vc{H}^{(i)}=\vc{q}_{i}\times\vc{E}^{(i)}$
(see Eq.~\eqref{eq:H-alp}).

As a result, we obtain 
the eigenvectors expressed as follows
\begin{align}
\label{eq:V_i_H_P_biax}
\vc{V}_i
\propto
&
\begin{pmatrix}
\vc{E}_{P}^{(i)}
\\
\vc{H}_{P}^{(i)}
\end{pmatrix},\quad
  \vc{H}_{P}^{(i)}=
\begin{pmatrix}
  q_{i} D_1^{(i)}\\
q_z^{(i)} D_2^{(i)}
\end{pmatrix},
\\
\label{eq:E_P_biax}
\mu^{-1} \vc{E}_{P}^{(i)}
=
&
\begin{pmatrix}
  q_z^{(i)} D_1^{(i)} /q_i  
\\
D_2^{(i)}
\end{pmatrix}
+q_x
\Bigl[
\Delta\eta_1
\sca{\vc{q}_i}{\uvc{d}}
\sca{\uvc{d}}{\vc{D}^{(i)}}
+
\notag
\\
&
\Delta\eta_2
\sca{\vc{q}_i}{\uvc{m}}
\sca{\uvc{m}}{\vc{D}^{(i)}}
\Bigr]
\begin{pmatrix}
1
\\
0
\end{pmatrix}.
\end{align}

\subsection{Uniaxial anisotropy}
\label{subsec:uniaxial-anisotropy}

Now we apply the above procedure to the limiting case
of uniaxial anisotropy with $\epsilon_2=\epsilon_3$
and $\Delta\epsilon_2=\Delta\eta_2=0$.
At $\Delta\eta_2=0$,
the Fresnel equation~\eqref{eq:eigval-qz}
takes the factorized form and 
the values of $q_z$ can be found as roots of
two quadratic equations.

The first equation $1-\eta_3 q^2=0$ represents the spherical wave
surface. The corresponding eigenmodes are known as 
the \textit{ordinary waves}.
There are two  values of $q_z$  
\begin{align}
  \label{eq:qz_pmo}
  q_{z}^{(\pm o)}=\pm\sqrt{n_{\perp}^2-q_x^2},
\end{align}
where $n_{\perp}^2=\mu\epsilon_3\equiv\mu\epsilon_{\perp}$,
that are equal in value but opposite in sign.
When, similar to the incident and transmitted waves, 
the $z$ component of the wave vector (and the vector $\vc{q}$)
is positive, the eigenmode might be called the 
\textit{refracted (forward) eigenwave}.
In the opposite case where, similar to the reflected wave, 
$q_{z}^{(\alpha)}$ is negative, the eigenmode will be referred to as 
the \textit{reflected (backward)  eigenwave}.
So, Eq.~\eqref{eq:qz_pmo} describes two ordinary eigenmodes:
the refracted eigenwave with $q_z=q_{z}^{(+o)}>0$ and the reflected
eigenwave with $q_z=q_{z}^{(- o)}<0$.

The second equation 
\begin{align}
  \label{eq:qze-eq}
  q^2+u_a\sca{\vc{q}}{\uvc{d}}^2-n_{\parallel}^2=0,
\end{align}
where
$n_{\parallel}^2=\mu\epsilon_1\equiv\mu\epsilon_{\parallel}$
and 
$u_a=-\Delta\eta_1/\eta_1=(n_{\parallel}^2-n_{\perp}^2)/n_{\perp}^2$
is the \textit{anisotropy parameter},
gives the values of $q_z$ for the eigenmodes known as 
the \textit{extraordinary waves}.
These are given by
\begin{align}
& 
 \label{eq:qz_pme}
  q_{z}^{(\pm e)}=[1+u_a d_z^2]^{-1}
\bigl\{
-u_{a} d_z d_x q_x\pm\sqrt{D}
\bigr\},
\\
&
\label{eq:discrim}
D = n_{\parallel}^2 
\bigl(1+u_a d_z^2\bigr) -
q_x^2 
\bigl[1+u_a (d_x^2+d_z^2)\bigr] 
\end{align}
where $d_x=\sca{\uvc{d}}{\uvc{x}}$ and
$d_z=\sca{\uvc{d}}{\uvc{z}}$.

At $u_a>0$ ($u_a<0$), in the $x$-$z$ plane, Eq.~\eqref{eq:qze-eq}
describes the ellipse with the major (minor) semi-axis of the length
$n_{\parallel}$ oriented perpendicular to
the projection of the director~\eqref{eq:director-d} on the plane of incidence
$(d_x,0,d_z)$. The length of minor (major) semi-axis, 
$\tilde{n}_{\perp}=[n_{\perp}^{-2}-u_a (d_y/n_{\parallel})^2]^{-1/2}$,
depends on the $y$ component of the director
and varies from $n_{\perp}$ to $n_{\parallel}$
as $d_y^2$ increases from zero to unity.
Clearly, degeneracy in refractive indices with $n_o=n_{\pm e}$ 
may occur only if the director is in the incidence plane
($\phi_{\dd}=0$). Additionally,
the matching condition for the $x$ components
of $\vc{q}$ and the director 
$q_x\equiv n_m\sin\theta_{\inc}=\pm n_{o} d_x\equiv
\pm n_o\sin\theta_{\dd}$ needs to be met.

The wave vectors and the refractive indices
of the normal modes are determined by the relation
\begin{equation}
  \label{eq:q_pm}
  \vc{q}_{\pm\alpha}= k_{\vac}^{-1}\vc{k}_{\pm\alpha}=
q_x\,\uvc{x}+q_{z}^{(\pm\alpha)}\,\uvc{z}=
n_{\pm\alpha}\uvc{k}_{\pm\alpha},
\quad \alpha\in\{o, e\},
\end{equation}
where $n_{\pm\alpha}=q_{\pm\alpha}$,
$n_{\pm o}=n_o=n_{\perp}$ is the ordinary refractive index
and
$n_{\pm e}$ is the refractive index of the extraordinary wave propagating
along the unit vector $\uvc{k}_{\pm e}$.

By substituting $\Delta\eta_2=0$ into
Eq.~\eqref{eq:D_i}
we can obtain the polarization vectors of
the electric displacement field~\cite{Kis:arxiv:2006,Kis:jpcm:2007}: 
\begin{subequations}
\label{eq:D_eig-vect}
\begin{align}
&
  \label{eq:D-ordn}
  \vc{D}^{(\pm o)}
=
-q_{\pm o}^2 
\vc{q}_{\pm o}\times\uvc{d}
=
-n_{\perp}^2 
\vc{q}_{\pm o}\times\uvc{d},
\\
&
\label{eq:D-extr}
    \vc{D}^{(\pm e)}=
q_{\pm e}^2 \mvc{P}(\uvc{k}_{\pm e})\cdot\uvc{d}
=
q_{\pm e}^{\,2}\,\uvc{d} -\sca{\uvc{d}}{\vc{q}_{\pm e}}\vc{q}_{\pm e}.
\end{align}
\end{subequations}
Then, following the procedure described at the end of
the preceding section,
we find the polarization vectors of the electric field
for the eigenmodes 
\begin{subequations}
\label{eq:E_eig-vect}
\begin{align}
&
  \label{eq:E-ordn}
\mu^{-1}\, 
  \vc{E}^{(\pm o)}
=-\vc{q}_{\pm o}\times\uvc{d},
\\
&
  \label{eq:E-extr}
 \mu^{-1}\, 
 \vc{E}^{(\pm e)}=
\uvc{d} -
n_{\perp}^{-2}
\sca{\uvc{d}}{\vc{q}_{\pm e}}\vc{q}_{\pm e}.
\end{align}
\end{subequations}

The result for the magnetic field of the normal modes is
\begin{subequations}
\label{eq:H_eig-vect}
\begin{align}
&
\label{eq:H-ordn}
\vc{H}^{(\pm o)}=
n_{\perp}^{2}\,\uvc{d}- 
\sca{\uvc{d}}{\vc{q}_{\pm o}}\vc{q}_{\pm o},
\\
&
\label{eq:H-extr}
  \vc{H}^{(\pm e)}=
\vc{q}_{\pm e}\times\uvc{d}.
\end{align}
\end{subequations}

Solution of the eigenvalue problem 
can now be presented in the matrix form:
\begin{align}
  \label{eq:V-nlc}
\bs{\Lambda}=
\diag(q_z^{(+e)},q_z^{(+o)},q_z^{(-e)},q_z^{(-o)}),
\quad
  \mvc{V}=
\begin{pmatrix}
\mvc{E}_{+} & \mvc{E}_{-}\\
  \mvc{H}_{+}&  \mvc{H}_{-}
\end{pmatrix},
\end{align}
where
\begin{align}
  \label{eq:E_pm}
  \mvc{E}_{\pm}=
\begin{pmatrix}
E^{(\pm e)}_x & E^{(\pm o)}_x\\
E^{(\pm e)}_y & E^{(\pm o)}_y
\end{pmatrix},
\quad
  \mvc{H}_{\pm}=
\begin{pmatrix}
H^{(\pm e)}_y & H^{(\pm o)}_y\\
-H^{(\pm e)}_x & -H^{(\pm o)}_x
\end{pmatrix}.
\end{align}

Owing to the orthogonality relations~\eqref{eq:orth_rel}
the matrix 
\begin{align}
  \label{eq:N_nlc}
 \mvc{N}=\tcnj{\mvc{V}}\cdot\mvc{G}\cdot\mvc{V}=\diag(\mvc{N}_{+},\mvc{N}_{-}) 
\end{align}
is diagonal and its non-vanishing elements
\begin{align}
  \label{eq:N_pm_el}
  \mvc{N}_{\pm}=
\begin{pmatrix}
N_{\pm e} & 0\\ 0& N_{\pm o}
\end{pmatrix},
\quad N_{\alpha} =
 2 \sca{\uvc{z}}{\vc{E}^{(\alpha)}\times\vc{H}^{(\alpha)}}
=
2 \sca{\vc{E}_{P}^{(\alpha)}}{\vc{H}_{P}^{(\alpha)}}
\end{align}
are proportional to the normal components of the Poynting vector
of the eigenmodes.

Substituting the expressions for the electric and magnetic
fields of the eigenmodes 
[see Eq.~\eqref{eq:E_eig-vect} 
and Eq.~\eqref{eq:H_eig-vect}, respectively]
into Eq.~\eqref{eq:N_pm_el} gives
the diagonal elements of the matrices 
$\mvc{N}_{+}$ and $\mvc{N}_{-}$
\begin{align}
&
  \label{eq:N_pmo}
  \mu^{-1}\,N_{\pm o}=
2 q_z^{(\pm o)}
\bigl[
n_{\perp}^2 - \sca{\vc{q}_{\pm o}}{\uvc{d}}^2
\bigr],
\\
&
  \label{eq:N_pme}
  \mu^{-1}\,N_{\pm e}=
2 n_{\perp}^{-2}
\Bigl\{
q_z^{(\pm e)}
\bigl[n_{\perp}^2 -
\sca{\vc{q}_{\pm e}}{\uvc{d}}^2
\bigr]
+
d_z \sca{\vc{q}_{\pm e}}{\uvc{d}}
\bigl[
n_{\pm e}^2 - n_{\perp}^2
\bigr]
\Bigr\}.
\end{align}

The eigenvalues
of ordinary and extraordinary waves
that enter the matrix of eigenvalues, $\bs{\Lambda}$, 
are given by Eq.~\eqref{eq:qz_pmo} and Eq.~\eqref{eq:qz_pme},
respectively.
The elements of the $2\times 2$ block matrices~\eqref{eq:E_pm}
that define the matrix of eigenvectors~\eqref{eq:V-nlc}
are the lateral components of the vectors~\eqref{eq:E_eig-vect}
and~\eqref{eq:H_eig-vect}.

These analytical expressions  
can now be substituted into 
the linking matrix~\eqref{eq:W-uni}
to write explicitly the elements of
the transfer matrix~\eqref{eq:trans-mat1}. 
For the homeotropic structure with
$\uvc{d}=\uvc{z}$,
it is rather straightforward to show that
the result is given by Eqs.~\eqref{eq:eigv_homeot}--\eqref{eq:gamma-homeotr}.


\end{document}